\ifdefined\pdfoutput\pdfoutput=1\fi
\documentclass[11pt]{article}

\usepackage[margin=1in]{geometry}
\usepackage{amsmath,amssymb,booktabs,array,microtype,graphicx,float}
\usepackage{xcolor,color}
\usepackage{iftex}
\ifPDFTeX
\usepackage[T1]{fontenc}
\usepackage{lmodern}
\else
\usepackage{fontspec}
\usepackage{unicode-math}
\fi
\usepackage{forest}
\useforestlibrary{edges}
\usepackage[colorlinks=true,citecolor=blue,linkcolor=black,urlcolor=black]{hyperref}

\title{Replica Fragmentation and Glassy Dynamics\\
in Parity Learning}
\author{Han Ma\\
\small CPHT, CNRS, École Polytechnique, Institut Polytechnique de Paris,\\
\small Palaiseau 91128, France}
\date{}
\hypersetup{
pdftitle={Replica Fragmentation and Glassy Dynamics in Parity Learning},
pdfauthor={Han Ma}
}

\newcommand{\E}{\mathbb{E}}

\newcommand{\sigmahat}{\widehat{\sigma}}

\newcommand{\dfnd}[1]{#1}
\newcommand{\pcmk}[1]{#1}

\begin{document}
\maketitle

\begin{abstract}
We study how independently trained Transformer neural networks reconstruct a binary string from its
local domain walls. Runs sharing the data and training protocol can realize different functions.
We treat them as replicas and measure truth alignment $m$, prediction confidence
$q_{\mathrm{self}}$, and cross-replica agreement $q_{\mathrm{cross}}$. Confident disagreement
defines the finite-size replica fragmentation that we call glass-like.
With small training sets, replicas predict all training examples correctly but remain confident
in incorrect predictions for unseen inputs, a regime we call memorization.
With larger sets, runs can generalize and then retreat. Retreat occurs when outputs start to deviate from 
truth while confidence remains high. The frontier between learned and
unlearned outputs recedes toward shorter strings. Many later recover as the frontier advances again.
The self--cross gap $q_{\mathrm{self}}-q_{\mathrm{cross}}$ clearly distinguishes the three
learning regimes of memorization, retreat, and recovery. With overall
and position-dependent truth alignment subtracted, their residual correlations also differ\pcmk{:} memorizing
replicas have weakly and uniformly correlated residuals, while retreat has the largest fraction of replica pairs 
whose residuals are anti-correlated. That is, on the inputs where one replica of such a pair does 
better than its average, the other tends to do worse. These finite-size
observations distinguish persistent memorization from ongoing retreat--recovery dynamics.
The thermodynamic and long-time limits, and extensions to other learning tasks, remain open.

\end{abstract}

\section{Introduction}
\label{sec:intro}

Machine learning emphasizes predictive accuracy and generalization
\cite{mehta2019high,carleo2019machine}, but repeated applications of a
fixed training procedure need not select the same predictive function. We consider
independent runs that share the training set and deterministic protocol but
vary in initialization and optimization noise. Such runs may attain similar
accuracy while differing in confidence and individual predictions
\cite{geiger2020scaling}. We study whether they select a functionally unique solution,
defined by agreement on a common set of held-out inputs.

These runs serve as replicas, and their learned functions form an ensemble.
Three overlaps characterize the ensemble\pcmk{:} truth alignment $m$, confidence
$q_{\mathrm{self}}$ (the mean squared soft spin), and agreement between independent replicas
$q_{\mathrm{cross}}$. Their differences define the Nishimori gap $\Delta_N$ and the self--cross gap
$\chi_{\mathrm{SG}}$\pcmk{:}
\begin{align}
\Delta_N &= q_{\mathrm{self}} - m,
\label{eq:Delta-N}\\
\chi_{\mathrm{SG}} &= q_{\mathrm{self}} - q_{\mathrm{cross}}.
\label{eq:chi-SG}
\end{align}
Here $\chi_{\mathrm{SG}}$ denotes the prediction variance across replicas. It measures
fragmentation of the learned-function ensemble and provides a finite-size overlap quantity whose
collective behavior can be tested through system-size scaling~\cite{mezard1987spin}. A positive
$\Delta_N$ measures confidence in excess of truth alignment.
The equality $q_{\mathrm{self}}=m$ holds on the Nishimori line for replicas drawn from a
posterior matched to the data-generating process
\cite{iba1999nishimori,nishimori2001statistical,zdeborova2016statistical}.
\dfnd{Our replicas are states recorded during optimization, so $q_{\mathrm{self}}\approx m$ is an empirical
relation between measured moments.}

For decades, overlap observables have provided a natural language for describing
memory retrieval and glassy organization in Hopfield's attractor model and its statistical-mechanical
extensions \cite{hopfield1982neural,amit1985storing,amit1987statistical}. Retrieval
corresponds to macroscopic alignment between a network state and a stored pattern,
whereas a state may be frozen without aligning with any particular memory. The statistical mechanics of learning has long borrowed this
machinery~\cite{engel2001statistical}. Loss surfaces and
weight-space training dynamics~\cite{baityjesi2018comparing} have both been compared
to glassy systems, and the dynamical comparison found no persistent aging in typical supervised
setups.
We apply these ideas to learned functions, with the target as a planted reference.
We call a shared-training-set ensemble glass-like when its replicas make confident but
distinct predictions. \dfnd{This interpretation rests on their overlap structure and dynamics at
finite size, without asserting a thermodynamic phase.}

To test this framework, we learn binary string reconstruction using neural networks. The
task has a unique target, and its position-resolved outputs expose partial learning.
Given a fixed boundary spin, the intervening domain walls determine a distant spin.
Equivalently the task computes an open Pauli string or \(Z_2\) Wilson line from the
local domain-wall variables. This local-to-global structure appears in stabilizer
quantum error correction, where local check measurements are used to infer a recovery
that preserves encoded logical
information~\cite{dennis2002topological,torlai2017neural}. It also appears in monitored
systems with a conserved charge, where an observer uses a record of local projective
measurements to infer the global charge of the initial state
\cite{barratt2022transitions}. Applying the \(m\)--\(q\) framework to this learning
problem via Transformers \cite{vaswani2017attention}, we observe two distinct
finite-size glass-like regimes.

For this position-resolved task, we write each learned function schematically as
\begin{equation*}
\begin{aligned}
\textit{learned function relative to the target}
={}&\textit{overall alignment}+\textit{position-resolved alignment}\\
&+\textit{shared residual}+\textit{individual residual}.
\end{aligned}
\end{equation*}
\noindent
The first two terms describe mean alignment across all outputs and at each position.
The residual splits into a part common to all replicas and a replica-dependent part
(Eqs.~\eqref{eq:per-replica-profile-decomposition}
and~\eqref{eq:shared-replica-residual-decomposition}). The common part cancels from the
self--cross gap but contributes to residual similarities.

At small training-set sizes, replicas fit every training example correctly, namely interpolate,
but make distinct,
confident predictions on held-out inputs, with weakly and uniformly correlated residual functions.

At larger training-set sizes, some runs first generalize and then lose much of
that performance under continued training, a phenomenon we call retreat.
Individual runs can undergo repeated retreat--recovery episodes. Related
observations include nonmonotonic test error, loss of acquired
generalization, and cyclic late-training dynamics under adaptive optimizers
\cite{advani2020high,he2024learning,thilak2022slingshot}.
\dfnd{Here training and held-out accuracy improve at approximately the same time
before the reversals, unlike canonical grokking, where generalization follows a
prolonged period of fitting only the training data.}
In addition to the accuracy studied in these works, we track the overlaps.
At larger training-set sizes, confidence is close to truth alignment, yet the
replicas still realize distinct functions.

Sec.~\ref{sec:binary-functions} defines the task, protocol and observables.
Sec.~\ref{sec:generalization} follows learning and retreat--recovery dynamics\pcmk{;}
Sec.~\ref{sec:classical-glasses} compares the residual geometry of memorization, retreat and
recovery. Sec.~\ref{sec:discussion} discusses interventions and limitations.
In a companion paper, we formulate the same task within a quantum framework and analyze
the learned functions with quantum-information quantities~\cite{ma2026percolation}.

\section{String reconstruction and learning regimes}
\label{sec:binary-functions}

To compare the functions selected by independent training runs, we 
define a string-reconstruction task with a unique target at each output position 
and specify the shared training and evaluation protocol. 

\subsection{The string-reconstruction task}
\label{subsec:binary-string-reconstruction}

Consider a binary domain-wall configuration
$b=(b_1,\ldots,b_D)\in\{0,1\}^D$. There are $L=D+1$ outputs, indexed by
$j\in\{0,\ldots,D\}$. Each binary target is the parity of the
first $j$ bits,
\begin{equation}
y_j^{\mathrm{bit}}(b)=\bigoplus_{i=1}^{j}b_i,
\qquad y_0^{\mathrm{bit}}=0.
\label{eq:intro-string-xor}
\end{equation}
Here $\oplus$ denotes addition modulo two.
For the same domain-wall configuration, introduce the spin representation $s=s(b)$, with
$s_i(b)=(-1)^{b_i}$. The
corresponding signed target is
\begin{equation}
\sigma_j^{\mathrm{true}}(s)=(-1)^{y_j^{\mathrm{bit}}(b)}
=\prod_{i=1}^{j}s_i\equiv\chi_{[1,j]}(s),
\qquad \sigma_0^{\mathrm{true}}=1.
\label{eq:signed-target-character}
\end{equation}
Thus $y_j^{\mathrm{bit}}$ and $\sigma_j^{\mathrm{true}}$ are the $\{0,1\}$ and $\{-1,+1\}$
representations of the same target. Each replica $r$ indexes an independent training run and produces
a soft spin
\begin{equation}
\sigmahat_{r,j}(b)\in[-1,1],
\end{equation}
where $\sigmahat_{r,j}=0$ corresponds to equal probability on the two classes, and
$|\sigmahat_{r,j}|\approx1$ indicates high confidence. Sec.~\ref{subsec:transformer}
gives the explicit $\tanh$ mapping and the shared-training-set protocol.

The target at $j$ is a degree-$j$ Boolean function and a Walsh character given by a product of input spins on
$\{0,1\}^D$~\cite{odonnell2014analysis}. Parity functions also form the Boolean
Fourier basis used to analyze the reconstruction and learning complexity of
many-body wavefunction signs~\cite{schurov2025learning}. This connects the present
task to sign-generalization difficulties found in neural representations of
frustrated magnets~\cite{westerhout2020generalization}. These targets also mirror error correction.
For a repetition-code chain with physical error bits
$e_i$, the local syndrome is
\begin{equation}
b_i=e_i\oplus e_{i+1}.
\end{equation}
After fixing the boundary bit $e_1$, recovery is syndrome integration,
\begin{equation}
e_{j+1}=e_1\oplus b_1\oplus\cdots\oplus b_j.
\end{equation}
String reconstruction is thus syndrome integration with a fixed boundary.
The output coordinate \(j\) measures both string length and algebraic degree, showing 
how far along the string successful reconstruction extends.
For replica $r$, the accuracy at output position $j$ and the block accuracy are
\begin{align}
A_{r,j}&=\E_b\!\left[\mathbb{I}\!\left(\operatorname{sgn}\sigmahat_{r,j}(b)
=\sigma_j^{\mathrm{true}}(s(b))\right)\right],
\label{eq:position-accuracy}\\
A_{r,\mathrm{block}}&=\E_b\!\left[\prod_{j=0}^{D}\mathbb{I}\!\left(
\operatorname{sgn}\sigmahat_{r,j}(b)=\sigma_j^{\mathrm{true}}(s(b))\right)\right],
\label{eq:block-accuracy}
\end{align}
where $\mathbb{I}$ is the indicator function and $\E_b$ averages over the evaluation set.
Eq.~\eqref{eq:block-accuracy} counts a configuration only when all $L$ output positions are correct
at once. We suppress the replica index where a single run is meant.

When truth alignment crosses from strong at small $j$ to weak at larger $j$, we call
the transition region the learning frontier and the weakly aligned or uncertain region beyond
it the tail.

For binary soft spins the overlaps of Sec.~\ref{sec:intro} are scalar products. We define the
truth overlap, self-overlap, and cross-replica overlap by
\begin{align}
m &= \E_{r,b,j}\left[\sigmahat_{r,j}(b)\sigma_j^{\mathrm{true}}(s(b))\right],
\label{eq:binary-truth-overlap}\\
q_{\mathrm{self}} &= \E_{r,b,j}\left[\sigmahat_{r,j}(b)^2\right],
\label{eq:binary-self-overlap}\\
q_{\mathrm{cross}} &= \E_{r\neq r',b,j}\left[\sigmahat_{r,j}(b)\sigmahat_{r',j}(b)\right].
\label{eq:binary-cross-overlap}
\end{align}
Here $\E_{r,b,j}$ denotes averaging over replicas $r$, domain-wall configurations $b$, and
output positions $j$.
The soft spin plays the role of a local magnetization, making $q_{\mathrm{self}}$ an
Edwards--Anderson-like self-overlap that measures confidence at a saved training state, also 
called checkpoint.
Temporal persistence is measured separately through two-time overlaps in
Sec.~\ref{sec:small-data-glass}. With $R$ replicas, the restriction $r\neq r'$ in
Eq.~\eqref{eq:binary-cross-overlap} makes $\chi_{\mathrm{SG}}$ exactly the variance of the
predictions across replicas,
\begin{equation}
\chi_{\mathrm{SG}}
=\E_{b,j}\!\left[\frac{1}{R-1}\sum_{r=1}^{R}
\left(\sigmahat_{r,j}(b)-\overline{\sigmahat}_{j}(b)\right)^2\right],
\qquad
\overline{\sigmahat}_{j}(b)=\frac{1}{R}\sum_{r=1}^{R}\sigmahat_{r,j}(b).
\label{eq:chi-replica-variance}
\end{equation}

\subsection{Learning regimes in the $m$--$q$ coordinates}

Table~\ref{tab:binary-mq-regimes} classifies five learning regimes by their overlaps and gaps.

\begin{table}[!htbp]
\centering
\small
\begin{tabular}{@{}>{\raggedright\arraybackslash}m{4.4cm}cccccc@{}}
\toprule
Learning class & $m$ & $q_{\mathrm{self}}$ & $q_{\mathrm{cross}}$ & $\Delta_N$ & $\chi_{\mathrm{SG}}$ & Physical analogue \\ \midrule
Retrieval & $\simeq1$ & $\simeq1$ & $\simeq1$ & $\simeq0$ & $\simeq0$ & Ordered-like \\
Uncertain & $\simeq0$ & $\simeq0$ & $\simeq0$ & $\simeq0$ & $\simeq0$ & Paramagnetic-like \\
Systematic error & $\ll1$ & $\simeq1$ & $\simeq1$ & $>0$ & $\simeq0$ & Spurious retrieval \\
Fragmented non-retrieval\newline(memorization) & $\simeq0$ & $\simeq1$ & $<q_{\mathrm{self}}$ & $>0$ & $>0$ & Spin-glass-like \\
Fragmented partial-retrieval\newline(retreat) & $0<m<1$ &
$O(1)$ & $<q_{\mathrm{self}}$ &
--- & $>0$ & Mixed order and glass \\ \bottomrule
\end{tabular}
\caption{Learning regimes relevant to the string-reconstruction task and
their statistical-physics analogues. A dash marks a quantity left
unconstrained by the state definition. \dfnd{These analogues are operational and finite-size
rather than thermodynamic phases.}}
\label{tab:binary-mq-regimes}
\end{table}

We focus on the last two, fragmented regimes. Memorization additionally requires
training-set interpolation, and retreat requires prior generalization\pcmk{;} the history-based
selection is specified in Sec.~\ref{sec:classical-glasses}.

\subsection{Transformer implementation and protocol}
\label{subsec:transformer}

The Transformer receives a two-channel input tensor $X(b)\in\mathbb{R}^{2\times L}$. The channels
place the adjacent domain-wall variables on the two sides of each output position, with missing boundary
values set to zero.
At each position, the two adjacent domain-wall values are first converted into an internal
vector, and a positional embedding encodes where that position lies along the chain. In
each layer, bidirectional self-attention allows every position to collect information from the complete
domain-wall configuration \cite{vaswani2017attention}. A feed-forward network (FFN) then processes the collected information separately at each
position. Layer normalization is applied before these two updates, and residual connections add each
update to the existing position representation \cite{ba2016layer,xiong2020layernorm,he2016residual}.
After the final layer, a linear readout converts the
representation at position $j$ into the logit $z_{r,j}(b)$, the log ratio of the predicted probabilities of bit 1 and bit 0. For replica $r$, we convert the logit
to the soft spin used in Sec.~\ref{subsec:binary-string-reconstruction},
\begin{equation}
\sigmahat_{r,j}(b)=-\tanh\left(z_{r,j}(b)/2\right)\in[-1,1].
\end{equation}
A positive logit favours $y_j^{\mathrm{bit}}=1$. The minus sign converts that convention into
the spin convention of Eq.~\eqref{eq:signed-target-character}.
Writing \(\boldsymbol{\vartheta}\) for the vector of trainable parameters, the scalar data
loss minimized on a minibatch \(\mathcal B\) of training examples is the mean binary cross-entropy,
\begin{equation}
\mathcal L_{\mathcal B}(\boldsymbol{\vartheta})
=\frac{1}{|\mathcal B|L}
\sum_{(b,y)\in\mathcal B}\sum_{j=0}^{L-1}
\ell_{\mathrm{BCE}}\!\left(z_{\boldsymbol{\vartheta},j}(b),
y_j^{\mathrm{bit}}(b)\right).
\label{eq:training-bce}
\end{equation}
Here $\ell_{\mathrm{BCE}}$ is the negative log-probability assigned to the correct binary label.
The optimizer AdamW applies weight decay directly to the parameters rather than adding an \(L_2\) penalty to the data loss.

The network carries $2.7\times10^{4}$ trainable parameters in every experiment below, while the
number of supervised bits $N_{\mathrm{train}}\times L$ ranges from $768$ to $4.9\times10^{5}$.
The small-data
ensembles are overparametrized by factors of $6$--$35$ and empirically reach exact interpolation,
whereas the largest retreat ensemble, at $(20,24576)$, contains eighteen times more supervised
bits than parameters. Across the parameter-to-data ratios studied, fragmentation is not tied to overparameterization\pcmk{:} it appears both in the overparameterized small-data ensembles and in the largest retreat ensemble, which has more supervised bits than parameters.

In the shared-training-set replica protocol, runs share the training set, architecture,
optimizer, batch protocol and evaluation set\pcmk{;} initialization and stochastic training noise vary.
Earlier work has quantified the resulting variation in
decision boundaries~\cite{geiger2020scaling}.
Here we study the overlaps as observables which classify the learned functions.

The training set acts as fixed quenched disorder, and each stochastic training run selects
a state at the observation time shared within its ensemble. This resembles a maturation
metastate~\cite{newman1997metastate,white2006scenario}, which collects states over initial
conditions and dynamical histories at fixed disorder. \dfnd{We take neither its infinite-time nor
its infinite-volume limit\pcmk{;} our ensembles remain tied to the specified protocol and endpoint selection.}

Unless stated otherwise, the Transformer has eight encoder layers, a 16-component
representation at each position, four attention heads, feed-forward width 64,
Gaussian error linear unit (GELU) activations \cite{hendrycks2016gelu}, and no dropout. We train with
AdamW~\cite{loshchilov2019decoupled}, learning rate $2\times10^{-3}$, weight decay $10^{-4}$,
and batch size 128. One epoch is one pass through the fixed training set
and contains $\lceil N_{\mathrm{train}}/128\rceil$ parameter updates. $N_{\mathrm{train}}$ is the number of
draws rather than the number of distinct configurations, since the draws are independent. The trajectory analyses of Sec.~\ref{sec:generalization} and the
fragmentation analysis of Sec.~\ref{sec:classical-glasses} use different ensembles.
Tables~\ref{tab:trajectory-ensembles} and~\ref{tab:glass-training-ensembles} summarize
the trajectory ensembles and the fragmentation-analysis protocols, respectively.

\begin{table}[!htbp]
\centering
\small
\begin{tabular}{@{}llll@{}}
\toprule
Analysis & Replicas & Recorded along the trajectory & Evaluation configurations \\ \midrule
Secs.~\ref{subsec:timing-comparison}, \ref{subsec:retreat-recovery} & $16$ & overlaps at every epoch to $600$ & complete $2048$ \\
Sec.~\ref{subsec:spatial-retreat-recovery} & $16$ & soft outputs at every epoch to $150$ & $4096$ draws \\
Sec.~\ref{subsec:tail-sample-crossover} & $6$ per $N$ & endpoint at $\simeq4000$ updates & complete held-out \\ \bottomrule
\end{tabular}
\caption{Trajectory ensembles used in Sec.~\ref{sec:generalization}. The $4096$ draws are
uniform, so they repeat configurations, because $L=12$ admits only $2^{L-1}=2048$ distinct ones.}
\label{tab:trajectory-ensembles}
\end{table}

\begin{table}[!htbp]
\centering
\small
\begin{tabular}{@{}cccccc@{}}
\toprule
Regime & $(L,N_{\mathrm{train}})$ & Replicas & Endpoint
& Evaluation configurations & $\mathcal J_{\mathrm{tail}}$ \\ \midrule
Small data & $(12,64)$ & $100$ & $\simeq4000$ updates & complete held-out ($1984$) & $8{:}11$ \\
Small data & $(24,64)$ & $100$ & $\simeq4000$ updates & $4000$ & $18{:}23$ \\
Small data & $(24,128)$ & $100$ & $\simeq4000$ updates & $4000$ & $18{:}23$ \\
Small data & $(24,192)$ & $111$ & $\simeq4000$ updates & $4000$ & $18{:}23$ \\ \addlinespace
Large data (retreat) & $(12,1280)$ & $269$ & epoch $150$ & $2048$ & $8{:}11$ \\
Large data (retreat) & $(16,2048)$ & $96$ & epoch $150$ & $2048$ & $11{:}15$ \\
Large data (retreat) & $(16,2560)$ & $86$ & epoch $150$ & $2048$ & $11{:}15$ \\
Large data (retreat) & $(16,3072)$ & $106$ & epoch $150$ & $2048$ & $11{:}15$ \\
Large data (retreat) & $(20,24576)$ & $78$ & epoch $150$ & $2048$ & $14{:}19$ \\ \bottomrule
\end{tabular}
\caption{Fixed-training-set ensembles used in the fragmentation comparisons.
For $(24,192)$ we use only the replicas that interpolate the training set.
For the retreat ensembles we use only replicas that reached $A_{\mathrm{block}}\geq0.8$ earlier and
are below $0.5$ at epoch 150.
Evaluation configurations are held fixed across replicas.
The complete held-out set excludes the training set. The other evaluation sets are
independent draws and can repeat a training configuration. The notation $a{:}b$ denotes $j=a,\ldots,b$.}
\label{tab:glass-training-ensembles}
\end{table}

The small-data ensembles share a budget of minibatch updates, each one an AdamW step,
whereas the large-data ensembles share an endpoint epoch. \dfnd{Epoch 150 corresponds to $1500$
updates at $N_{\mathrm{train}}=1280$ and $28{,}800$ at $N_{\mathrm{train}}=24576$,
so the latter comparison does not fix the update count across training-set sizes.}

Two settings fall outside Table~\ref{tab:glass-training-ensembles}. For the $L=12$ training-set-size scan in
Fig.~\ref{fig:small-data-crossover} we evaluate on the complete set of domain-wall
configurations after removing the shared training set.
The convolutional neural network (CNN) used as an architecture control has 16 residual convolutional
layers, 16 channels, kernel size 3, GroupNorm with all 16 channels in a single group, GELU
activations, and a position-wise $1\times1$ readout \cite{he2016residual,wu2018group}.
In the $L=12$ CNN--Transformer comparison, both
models use the batch size and AdamW settings given above and are trained for 120 epochs.

\section{Generalization crossover and retreat dynamics}
\label{sec:generalization}

We first locate the generalization crossover and then follow the dynamics of retreat and recovery.
Previous work on modular arithmetic studied delayed generalization
and a critical training-set size in the long-training limit~\cite{gromov2023grokking}.
In the finite-budget scan of Sec.~\ref{sec:sample-controlled-generalization}, successful runs
cross the training and held-out thresholds at essentially the same epoch. Each run uses a
separately drawn fixed training set, so the learning probability averages over training-set
draws, initialization and training noise. By contrast, within each subsequent ensemble, all replicas
share a single fixed training set, so only initialization and training noise vary across replicas.
The evaluation set for each analysis is specified in Table~\ref{tab:trajectory-ensembles}.

\subsection{Generalization at large training-set sizes}
\label{sec:sample-controlled-generalization}

For each training-set size $N$, we count a run as successful if its held-out block
accuracy reaches $A_{\mathrm{block}}\geq0.8$ at least once during the 120-epoch scan. If $k_N$ of
the $n_N$ runs succeed, the measured finite-budget learning probability is
$\widehat P_{\mathrm{learn}}(N)=k_N/n_N$.\footnote{\label{fn:wilson}For a binomial run fraction
$\widehat p=k/n$, the Wilson score interval~\cite{wilson1927probable} is $[c-h,c+h]$, where
$c=[\widehat p+z^2/(2n)]/[1+z^2/n]$ and
$h=z[\widehat p(1-\widehat p)/n+z^2/(4n^2)]^{1/2}/[1+z^2/n]$.
We use $z=1$ here, corresponding to approximately 68\% coverage.
The interval quantifies run-level sampling uncertainty.}
We fit a descriptive logistic curve to the success and failure counts by binomial
maximum likelihood\pcmk{:}
\begin{equation}
P_{\mathrm{fit}}(N)
=\frac{1}{1+\exp\!\left[-(\log N-\log N_c)/w\right]}.
\label{eq:learning-probability-logistic}
\end{equation}
In Eq.~\eqref{eq:learning-probability-logistic}, $N_c$ is the fitted training-set
size at which $P_{\mathrm{fit}}=1/2$, and $w$
sets the crossover width along the $\log N$ axis. A smaller $w$ means a sharper crossover.
The fits
give $N_c\simeq349$ and $w=0.090$ for the CNN, compared with $N_c\simeq1111$ and $w=0.261$ for the
Transformer. Thus, under this protocol, the Transformer requires more training data
and has a broader crossover across runs. \dfnd{The across-run fit and the within-run
trajectory quantify different aspects of the crossover.} Variation in its location across
runs broadens the ensemble crossover, while individual successful Transformer runs can change
abruptly.

\begin{figure}[!htbp]
\centering
\includegraphics[width=\textwidth]{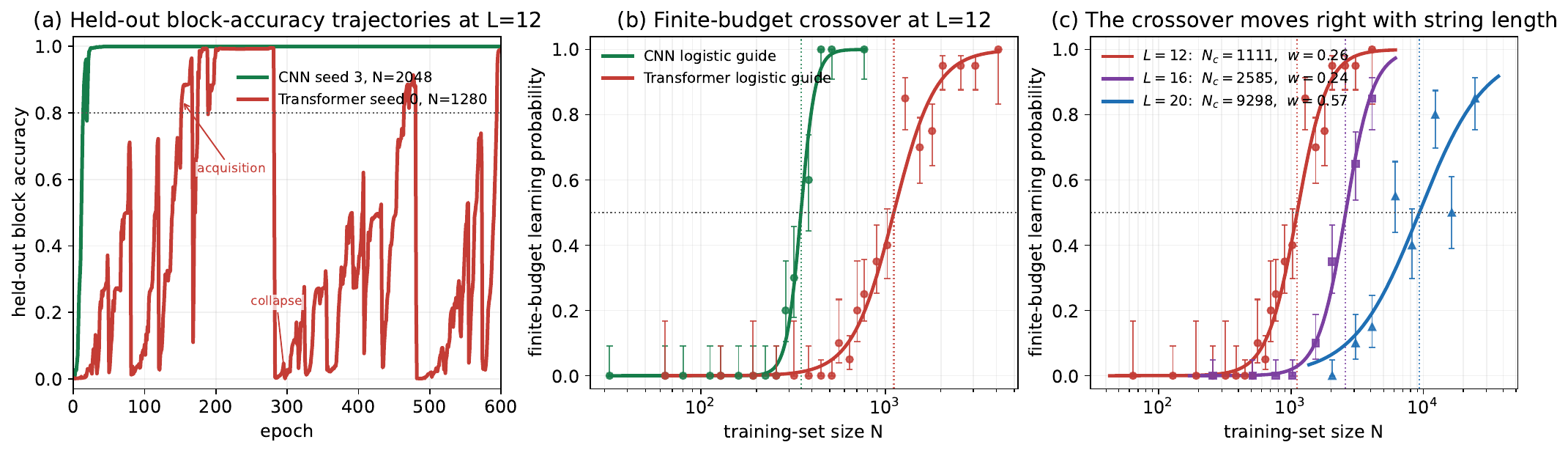}
\caption{Generalization with independently sampled fixed training sets. For
each training-set size $N$, every run samples its own training set from the domain-wall
distribution and holds it fixed throughout training. (a) Representative held-out block-accuracy trajectories for a
CNN at $N=2048$ and a Transformer at $N=1280$. The CNN improves progressively and remains accurate,
whereas the Transformer learns abruptly, later loses accuracy, and then recovers. (b) Fraction of
runs that reach $A_{\mathrm{block}}\geq0.8$ within 120 epochs, plotted against $N$. Points show the
measured fractions with binomial confidence intervals. Curves are the descriptive binomial-logistic
fits in Eq.~\eqref{eq:learning-probability-logistic}. (c) The same Transformer protocol at
$L=12$, $16$ and $20$, with the $L=12$ curve repeated from (b). Dotted vertical lines mark the
fitted $N_c$. The three curves do not cross inside the measured range.}
\label{fig:generalization-transition}
\end{figure}

The Transformer fit above gives $N_c(L=12)\simeq1.1\times10^3$ under the stated
$A_{\mathrm{block}}\geq0.8$ and 120-epoch criterion. Interpolating the same scan at $L=16$ and $L=20$ gives
$N_c(L=16)\simeq2.6\times10^3$ and $N_c(L=20)\simeq9.3\times10^3$
(Fig.~\ref{fig:generalization-transition}(c)). The crossover also broadens at the longest string\pcmk{:} the fitted width is
$w=0.26$ at $L=12$, $0.24$ at $L=16$ and $0.57$ at $L=20$. At $L=24$ we estimate the crossover near $3\times10^4$ without a
full scan, because the runs there are too expensive. \dfnd{This full-rule generalization
scale is distinct from the smaller crossover of Fig.~\ref{fig:small-data-crossover} below, which is
measured on the tail positions after a far longer training time at small $N$.}

The CNN control generalizes with fewer data and stays accurate after learning.
These distinct dynamics motivate a future comparison of the CNN and Transformer overlap geometries.
The replica analysis below focuses on Transformers.

Besides architecture, the sharp crossover of block accuracy in
Fig.~\ref{fig:generalization-transition} stems from its particular definition. The product over positions in
Eq.~\eqref{eq:block-accuracy} can make changes in performance appear sharper than changes in a
lower-level accuracy measure. Let $A_{\mathrm{bit}}$ be the mean of
Eq.~\eqref{eq:position-accuracy} over the $L$ output positions. If every position had accuracy
$A_{\mathrm{bit}}$ and failed independently, block accuracy would be $(A_{\mathrm{bit}})^{L}$.
Across the $L=12$ scan $A_{\mathrm{block}}$ and $(A_{\mathrm{bit}})^{12}$ have correlation $0.989$. The measured block accuracy is on average
$0.033$ higher than this estimate
(Appendix~\ref{app:accuracy-correlation-statistics}). The positive offset means correctness is
positively correlated across output positions. The predictions begin to improve before the block threshold is crossed.
Section~\ref{subsec:timing-comparison} compares their timing directly.

\subsection{Hidden progress before the block-accuracy crossing}
\label{subsec:timing-comparison}

\begin{figure}[!htbp]
\centering
\includegraphics[width=\textwidth]{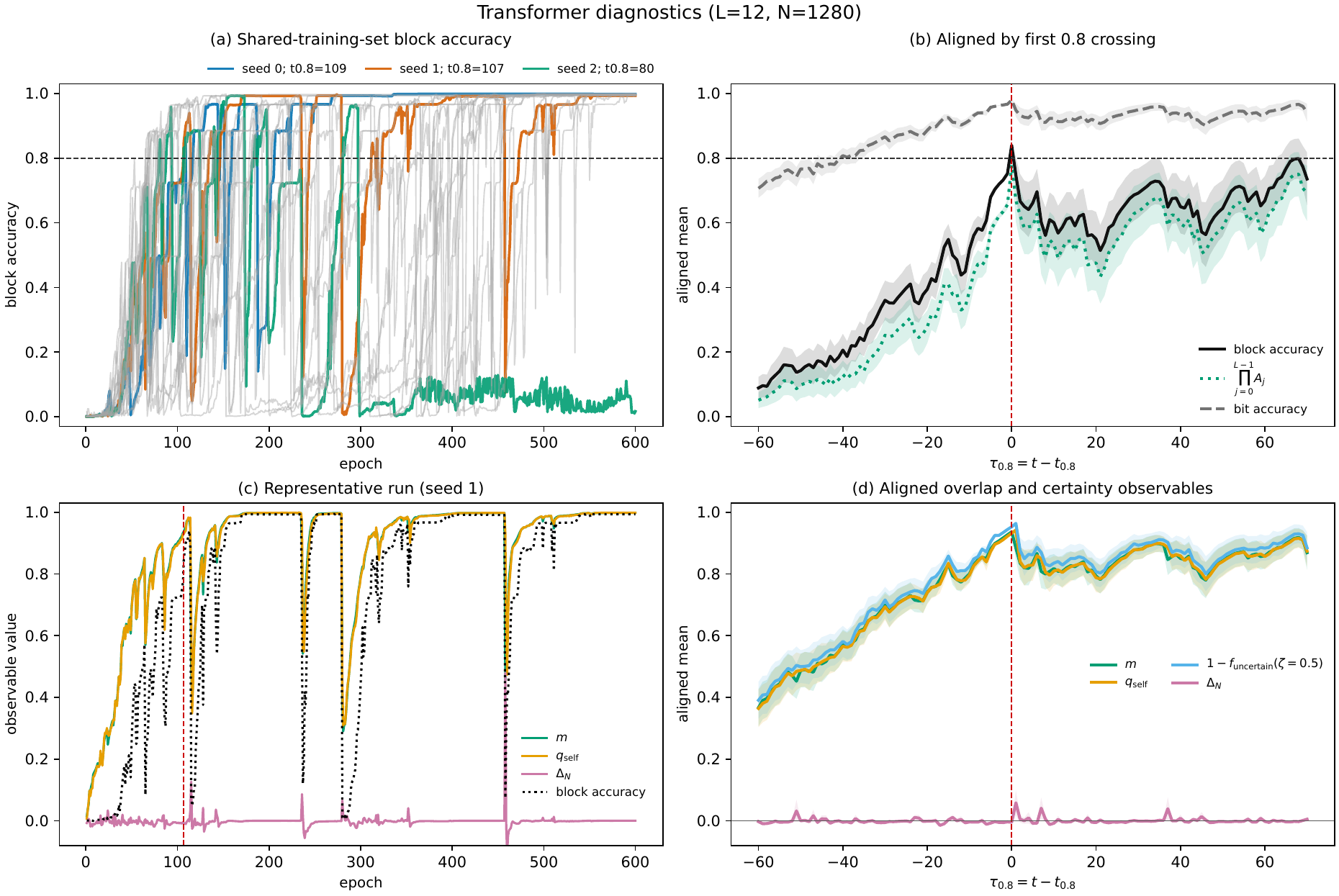}
\caption{Shared-training-set diagnostics over 600 epochs for $L=12$, $N=1280$
Transformer runs. The ensemble has 16 runs. The replicas share one fixed training set of 1280 configurations sampled
independently from the domain-wall distribution and are evaluated on the complete set of 2048
domain-wall configurations. Only initialization and minibatch order vary. (a) All
block-accuracy trajectories, with the pre-specified random seeds 0, 1, and 2 highlighted and the other
13 shown in gray. Every run undergoes multiple threshold crossings and a post-crossing retreat
below $0.5$. (b) Ensemble means after alignment by the first
$A_{\mathrm{block}}\geq0.8$ crossing. $A_{\mathrm{bit}}$ and $\prod_j A_j$ rise before the
block-accuracy threshold. The product $\prod_j A_j$ differs from $(A_{\mathrm{bit}})^{L}$, which
assigns the same mean accuracy to every position. (c) $m$,
$q_{\mathrm{self}}$, $\Delta_N$, and block accuracy for seed 1.
(d) Aligned ensemble means show $m$ and $q_{\mathrm{self}}$ building together before the
crossing, with sparse positive $\Delta_N$ excursions. Shaded bands are standard errors across the
runs.}
\label{fig:generalization-retreat}
\end{figure}

To resolve hidden progress and repeated threshold crossings, we follow a separate
shared-training-set Transformer ensemble at $L=12$, $N=1280$ for 600 epochs and evaluate
it on the complete input space (Fig.~\ref{fig:generalization-retreat}).

For run $r$, let $t_{0.8}^{(r)}$ denote the first epoch at which
$A_{\mathrm{block}}\geq0.8$, and define the aligned time
$\tau_{0.8}=t-t_{0.8}^{(r)}$. Panel (b) of
Fig.~\ref{fig:generalization-retreat} compares the measured block accuracy $A_{\mathrm{block}}$, the mean bit accuracy
$A_{\mathrm{bit}}$, and the product $\prod_{j=0}^{L-1}A_j$, which predicts block
accuracy if correctness at different positions is independent at the measured accuracies
$A_j$.
In these trajectories, $\prod_{j}A_j$ lies below the measured block accuracy near
$\tau_{0.8}=0$. Well before this point, block accuracy, $\prod_{j}A_j$, and $A_{\mathrm{bit}}$ all rise. 
We also define the uncertain-output fraction as
$f_{\mathrm{uncertain}}(\zeta)=\mathbb{E}_{r,b,j}[\mathbb{I}(|\sigmahat_{r,j}(b)|<\zeta)]$ and the
corresponding prediction certainty as $1-f_{\mathrm{uncertain}}(\zeta)$. Panels (c) and (d) compare these
quantities with $m$ and $q_{\mathrm{self}}$.

Model-internal diagnostics have revealed hidden progress in parity learning and grokking~\cite{barak2022hidden,gromov2023grokking}. Here $m$ and $q_{\mathrm{self}}$ reveal progress
directly in the outputs before the block-accuracy threshold is crossed.
In panel (a), all runs reach the threshold, with first crossings spanning epochs 65--337.
When the trajectories are aligned by their first crossings, $A_{\mathrm{bit}}$, truth overlap,
self-overlap, and prediction certainty are already increasing well before $\tau_{0.8}=0$.
\begin{samepage}
At $\tau_{0.8}=-20$, their means are
\begin{equation}
A_{\mathrm{bit}}=0.886,
\qquad m=0.741,
\qquad q_{\mathrm{self}}=0.740,
\qquad 1-f_{\mathrm{uncertain}}(0.5)=0.762.
\end{equation}
At the same aligned time, mean block accuracy is $0.394$ and
$\Delta_N=q_{\mathrm{self}}-m=-0.001$. Narrowing the uncertainty threshold to $\zeta=0.2$ raises the
mean certainty to $0.877$, so the high certainty at $\tau_{0.8}=-20$ is not an artifact of
the threshold choice. The mean Nishimori gap stays within $|\Delta_N|<0.015$ throughout
$-40\leq\tau_{0.8}\leq0$, showing that truth
alignment and confidence build together. The ensemble therefore carries hidden partial order
that block accuracy does not register.
\end{samepage}

\subsection{Generalization retreat and recovery}
\label{subsec:retreat-recovery}

\begin{figure}[!htbp]
\centering
\includegraphics[width=0.88\textwidth]{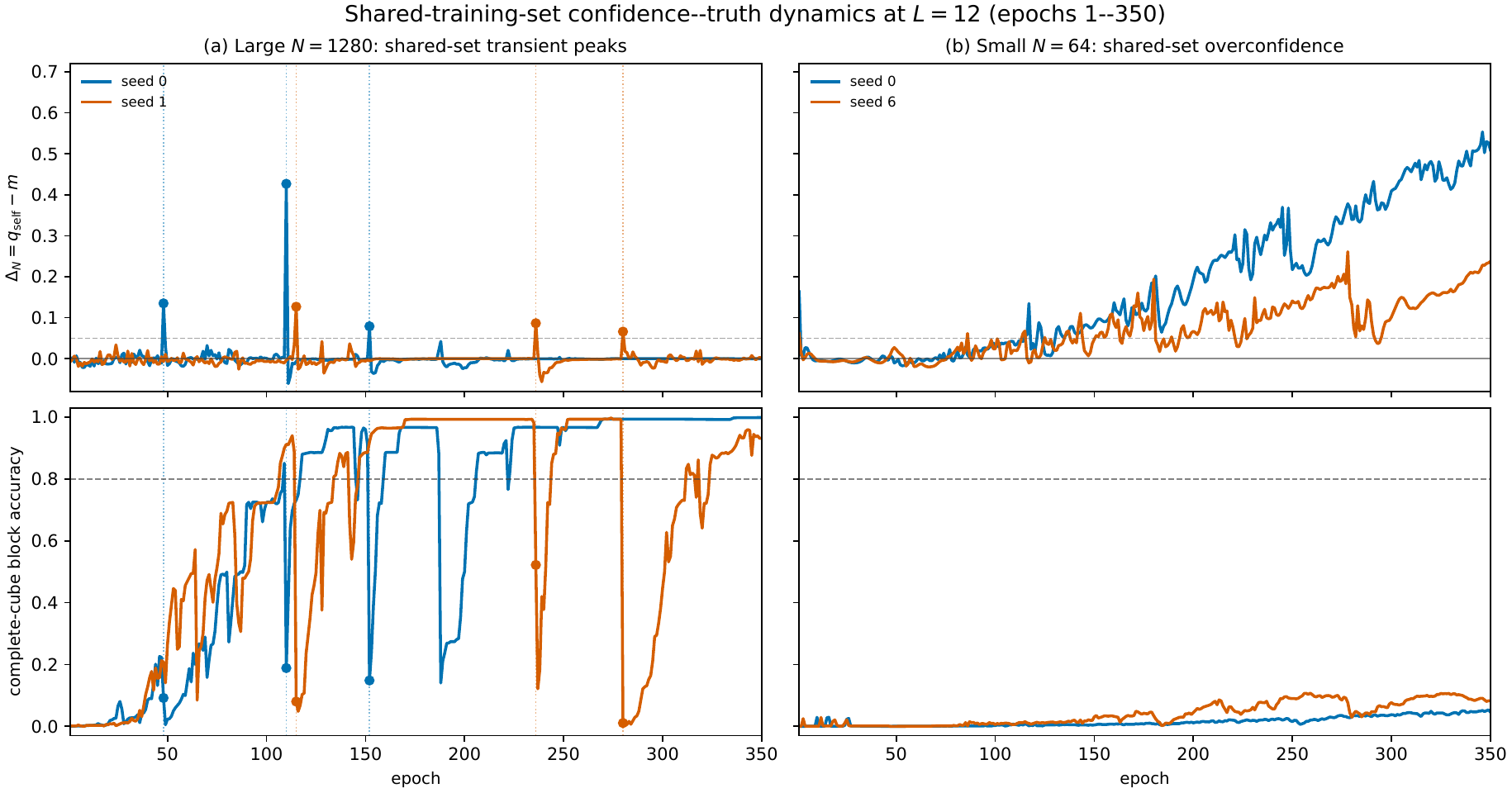}
\caption{Nishimori-gap dynamics evaluated on all 2048 domain-wall configurations at
$L=12$ over epochs 1--350. Left column\pcmk{:} the $N=1280$ runs with seeds 0 and 1, trained on the fixed
training set shared by all 16 replicas in Fig.~\ref{fig:generalization-retreat}. Dots mark local maxima of
$\Delta_N=q_{\mathrm{self}}-m$ above $0.05$, and vertical lines locate the same epochs in block
accuracy. Right column\pcmk{:} two independently initialized $N=64$ runs trained on the same fixed
training set\pcmk{;} positive $\Delta_N$ persists while block accuracy remains low.}
\label{fig:confidence-dynamics-comparison}
\end{figure}

After first reaching $A_{\mathrm{block}}\geq0.8$, every run falls below $0.5$
and later returns above $0.8$ at least once more. In panels (c) and (d) of Fig.~\ref{fig:generalization-retreat}, $m$ and
$q_{\mathrm{self}}$ usually continue to evolve together. At an abrupt retreat, however, $m$ can fall faster than
$q_{\mathrm{self}}$, producing a positive peak in
$\Delta_N=q_{\mathrm{self}}-m$. The gap then closes over the next few epochs as
$q_{\mathrm{self}}$ relaxes and $m$ recovers. Fig.~\ref{fig:confidence-dynamics-comparison}
emphasizes representative peaks and introduces a small-$N$ comparison where this
Nishimori gap is persistent.

We first ask how often the positive peaks characterizing large-$N$ retreat occur.
We record one checkpoint per epoch. Combining the epochs from each run's first
$A_{\mathrm{block}}\geq0.8$ crossing through epoch 600 gives 7915 replica--epoch observations,
and $98\%$ of them satisfy $|\Delta_N|\leq0.05$. Restricting to the observations with
$A_{\mathrm{block}}<0.5$ leaves $94\%$ within the same band. 
Thus, truth alignment and confidence remain closely
coupled during most large-$N$ retreats, and the positive peaks mark short-lived lags rather than a
sustained overconfident regime.

The small-$N$ trajectories show the opposite temporal pattern. In
Fig.~\ref{fig:confidence-dynamics-comparison}, both training-set sizes are evaluated on all 2048 domain-wall configurations
and displayed on the same vertical scale. The two $N=64$ runs provide 300 replica--epoch
observations over epochs 201--350, and $98\%$ of them have $\Delta_N>0.05$. Both gaps are
still large at epoch 350 while both block accuracies remain below $0.1$. Confidence therefore remains persistently separated from truth alignment at small $N$, in contrast
to the brief relaxation lag at large $N$.

Together, these results show what the joint $m$--$q_{\mathrm{self}}$ diagnosis adds
to descriptions of nonmonotonic accuracy
\cite{he2024learning}.
Retreat and memorization can both have low block accuracy, 
but their \(m\)–\(q_{\mathrm{self}}\) dynamics differ. Retreat shows brief 
confidence lags, whereas memorization sustains confident misalignment. 
Both moments average over output
positions. Appendix~\ref{app:retreat-curvature} examines the corresponding 
model-level signatures through evaluation-loss gradients, Hessian extrema, and 
target-specific structure in the final FFN. To determine whether retreat weakens 
all outputs uniformly or begins at particular
positions, we next resolve them along the chain.

\subsection{Position-resolved retreat and recovery}
\label{subsec:spatial-retreat-recovery}

\begin{figure}[!htbp]
\centering
\includegraphics[width=0.8\textwidth]{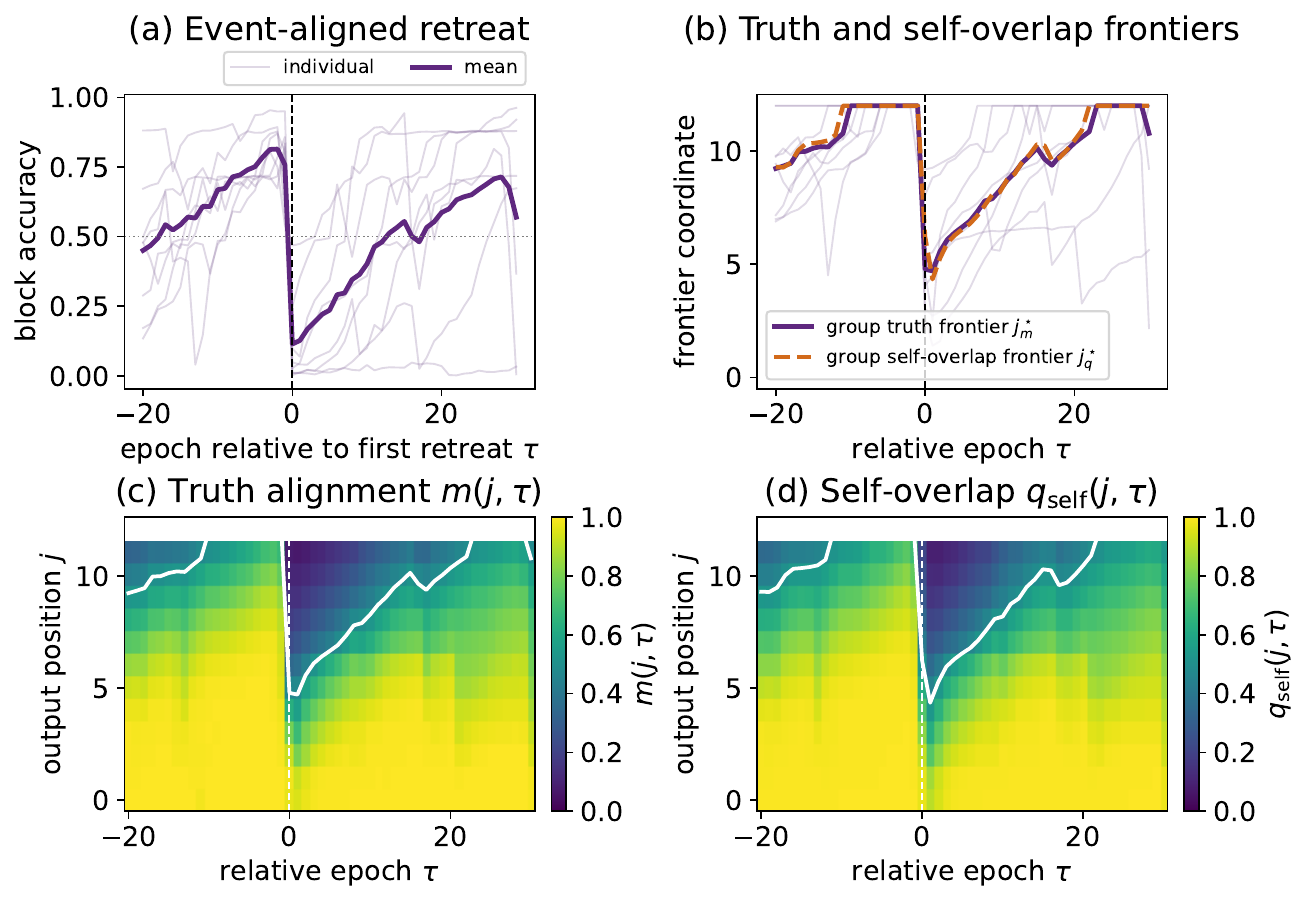}
\caption{Position-resolved retreat and recovery in the 16-replica shared-training-set
ensemble. The eight replicas that first reach $A_{\mathrm{block}}\geq0.8$ and later retreat below
$0.5$ are aligned at the first such retreat. (a) Individual and mean block accuracy. (b) The group frontiers $j_m^\star$ and
$j_q^\star$ with the individual truth frontiers $j_{m,r}^\star$ as faint curves. (c) and (d) Group profiles
$m(j,\tau)$ and $q_{\mathrm{self}}(j,\tau)$, both on a color scale running from $0$ to $1$. White
curves are the corresponding half-height frontiers, and the dashed vertical line in every panel marks
the retreat alignment point $\tau=0$. Frontiers are located by linear interpolation between
neighbouring positions. A profile entirely below one half is assigned $0$ and one entirely above is
assigned $L$.}
\label{fig:event-aligned-frontier-dynamics}
\end{figure}

By construction, output position indexes the target functions by string length and algebraic degree
(Sec.~\ref{subsec:binary-string-reconstruction}). Sequential learning of parities ordered by
degree is the staircase picture of parity learning~\cite{abbe2022merged}. 
The position-resolved overlap profiles distinguish uniform retreat, in which all positions
change together, from degree-ordered retreat, which moves along \(j\).
For this analysis, we record another ensemble on the same fixed training set,
as specified in the second row of Table~\ref{tab:trajectory-ensembles}.
Twelve replicas reach $A_{\mathrm{block}}\geq0.8$, and eight subsequently retreat below $0.5$.
Aligning those eight runs at their first post-generalization retreat defines $\tau=0$ and gives
the aligned trajectories in panel (a) of Fig.~\ref{fig:event-aligned-frontier-dynamics}.

For these aligned replicas, we define the per-replica truth-alignment and
self-overlap profiles by
\begin{equation}
m_r(j,\tau)=\E_b[\sigmahat_{r,j}(b,\tau)\sigma_j^{\mathrm{true}}(s(b))],
\qquad
q_{\mathrm{self},r}(j,\tau)=\E_b[\sigmahat_{r,j}(b,\tau)^2],
\label{eq:position-resolved-overlaps}
\end{equation}
with group averages $m(j,\tau)=\E_r[m_r(j,\tau)]$ and
$q_{\mathrm{self}}(j,\tau)=\E_r[q_{\mathrm{self},r}(j,\tau)]$.
Thus, $m(j,\tau)$ resolves truth alignment by output position, while
$q_{\mathrm{self}}(j,\tau)$ resolves prediction confidence without reference to the truth. The
individual truth frontier $j_{m,r}^\star(\tau)$, the group truth frontier $j_m^\star(\tau)$, and the
group self-overlap frontier $j_q^\star(\tau)$ are the output positions at which the corresponding
profile crosses one half, so that
$m_r(j_{m,r}^\star,\tau)=m(j_m^\star,\tau)=q_{\mathrm{self}}(j_q^\star,\tau)=1/2$.

Fig.~\ref{fig:event-aligned-frontier-dynamics} shows that the loss and recovery
of global truth alignment are spatially nonuniform. The profile in panel (c) shows that at retreat truth alignment is lost first at larger $j$,
while smaller-$j$ outputs remain aligned. In panel (b), at the onset of retreat, $j_q^\star$ exceeds
$j_m^\star$\pcmk{:} predictions remain confident at some larger-$j$ positions after losing truth alignment.
Immediately afterward, $j_q^\star$ falls below $j_m^\star$ while $j_m^\star$ stays nearly fixed.
Between $\tau=0$ and $\tau=1$ the self-overlap frontier drops from $6.2$ to $4.3$ and the truth
frontier moves only from $4.8$ to $4.7$. Both frontiers then turn upward at the same epoch and
advance in step. Truth alignment therefore returns at progressively later output
positions. Having resolved retreat and recovery in time, we next examine how
tail-averaged endpoint observables vary with training-set size.

\subsection{Tail crossover across training-set size}
\label{subsec:tail-sample-crossover}

We now vary the training-set size and measure the tail observables at the recorded endpoint.
We use the overlaps defined in
Eqs.~\eqref{eq:binary-truth-overlap}--\eqref{eq:binary-cross-overlap}, changing only the range of
the position average to $|\mathcal{J}_{\mathrm{tail}}|^{-1}\sum_{j\in\mathcal{J}_{\mathrm{tail}}}$,
where $\mathcal{J}_{\mathrm{tail}}=\{8,\ldots,11\}$ is the fixed tail window used in the ensemble analyses below. The restricted $q_{\mathrm{self}}$ and
$q_{\mathrm{cross}}$ are also used to compute $\chi_{\mathrm{SG}}$ through
Eq.~\eqref{eq:chi-SG}.
Errors are estimated by a leave-one-replica jackknife. Because $q_{\mathrm{cross}}$ and
$\chi_{\mathrm{SG}}$ average over pairs that share replicas, each replica is dropped in turn and
the spread of the six recomputed estimates gives the standard error.

\begin{figure}[H]
\centering
\includegraphics[width=0.95\textwidth]{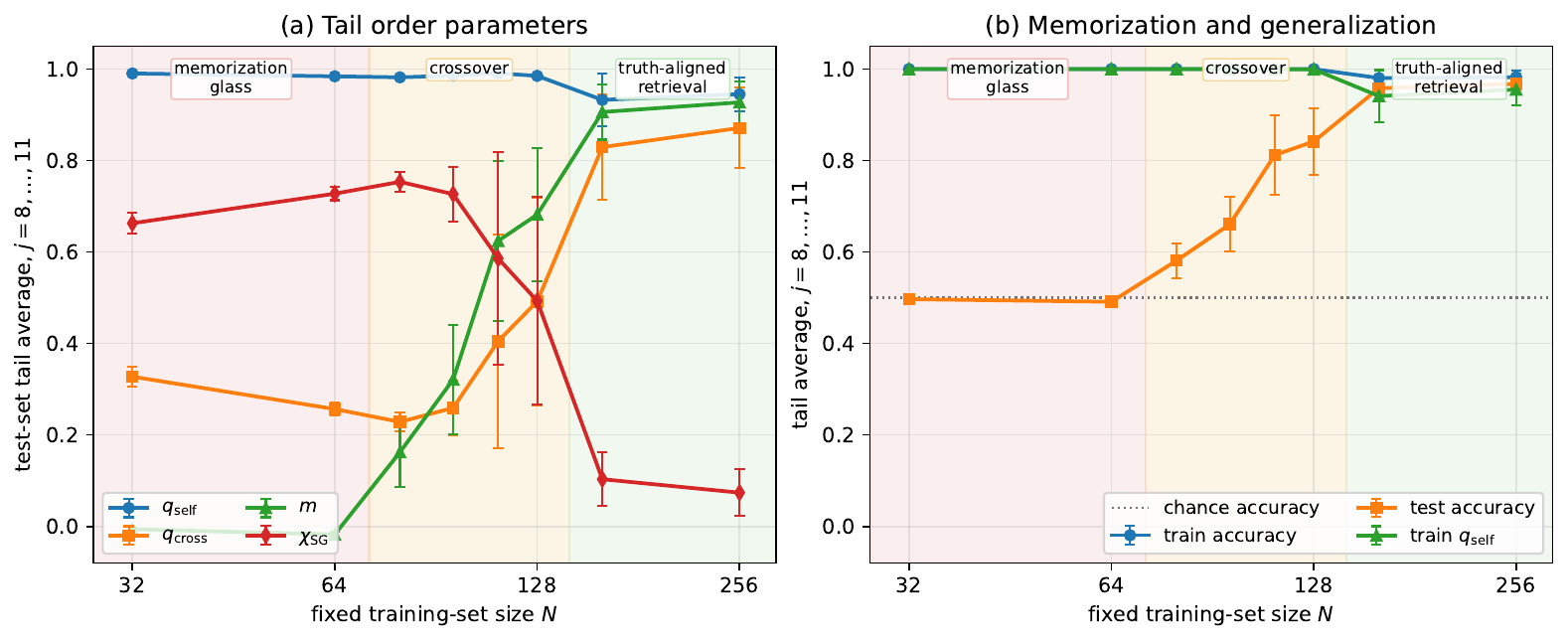}
\caption{Tail observables for $L=12$ versus fixed training-set size $N$.
(a) Truth overlap, self-overlap, cross-replica overlap, and the self--cross gap.
(b) Training and test accuracy together with training-set self-overlap. The dotted line is the
chance accuracy $0.5$ of a binary output. With batch size 128, all
endpoints use approximately 4000 updates\pcmk{:} 4000 epochs for $N\leq128$ and 2000 epochs for
$N=160,256$.
Error bars are leave-one-replica jackknife standard errors across six replicas.}
\label{fig:small-data-crossover}
\end{figure}

Fig.~\ref{fig:small-data-crossover} shows a crossover from confident memorization
($q_{\mathrm{self}}\simeq1$, $m\simeq0$ and a large self--cross gap) at $N\leq64$
to a shared, truth-aligned tail function by $N=256$. Training accuracy and training-set
self-overlap stay near one, while test accuracy rises from the binary chance level
$0.5$ for $N\leq64$ to nearly one by $N=256$.

\section{Geometry of the three learning regimes}
\label{sec:classical-glasses}

Sec.~\ref{sec:generalization} identified memorization and retreat--recovery dynamics.
We now compare their endpoint ensembles (Table~\ref{tab:glass-training-ensembles}), separating
the amount of fragmentation from its organization. After removing scalar truth alignment and
mean position profiles, we measure the residual gap and geometry, including departure from the
ultrametric hierarchy of a mean-field spin glass~\cite{mezard1987spin}.

\subsection{A decomposition of the self--cross gap}
\label{subsec:profile-residual-decomposition}

For replica $r$, evaluation configuration $b$, and output position $j$, the truth-aligned soft spin
is $u_{r,j}(b)=\sigmahat_{r,j}(b)\sigma_j^{\mathrm{true}}(s(b))$, so its sign indicates correctness
and its magnitude indicates confidence. The replica's mean truth-alignment profile and its position
average are
\begin{equation}
\mu_{r,j}=\E_b[u_{r,j}(b)],
\qquad
m_r=\E_j[\mu_{r,j}],
\end{equation}
where $\E_b$ averages over the common evaluation set and $\E_j$ over all output positions.
Writing $\delta\mu_{r,j}=\mu_{r,j}-m_r$ and
$\epsilon_{r,j}(b)=u_{r,j}(b)-\mu_{r,j}$ gives the per-replica decomposition
\begin{equation}
u_{r,j}(b)=m_r+\delta\mu_{r,j}+\epsilon_{r,j}(b).
\label{eq:per-replica-profile-decomposition}
\end{equation}
The three terms are the scalar truth alignment, the variation of the mean profile across
positions, and the input-dependent residual. By construction,
$\E_j[\delta\mu_{r,j}]=0$ for every replica and
$\E_b[\epsilon_{r,j}(b)]=0$ for every replica and position.
Collecting the scalars for $R$ selected replicas into
$\boldsymbol{m}=(m_1,\ldots,m_R)^\mathsf{T}$, the raw and connected overlaps are
\begin{equation}
Q_{r,s}^{\mathrm{raw}}=\E_{b,j}[u_{r,j}(b)u_{s,j}(b)],
\qquad
Q_{r,s}^{c}=Q_{r,s}^{\mathrm{raw}}-m_rm_s.
\label{eq:raw-connected-overlap}
\end{equation}
Because $[\sigma_j^{\mathrm{true}}(s(b))]^2=1$, the replica-averaged diagonal and
distinct-replica-averaged off-diagonal of $Q^{\mathrm{raw}}$ are $q_{\mathrm{self}}$ and
$q_{\mathrm{cross}}$.

The zero-mean conditions make the cross terms in
Eq.~\eqref{eq:per-replica-profile-decomposition} vanish under the joint average. The raw
overlap therefore decomposes as
\begin{equation}
Q^{\mathrm{raw}}=\boldsymbol{m}\boldsymbol{m}^\mathsf{T}+C^{\mathrm{prof}}+W,
\label{eq:raw-profile-decomposition}
\end{equation}
\begin{align}
C_{r,s}^{\mathrm{prof}}
&=\operatorname{Cov}_j(\mu_{r,j},\mu_{s,j})
{}=\E_j[\mu_{r,j}\mu_{s,j}]-\E_j[\mu_{r,j}]\,\E_j[\mu_{s,j}],
\label{eq:profile-covariance} \\
W_{r,s}
&=\E_j\!\left[\operatorname{Cov}_b(u_{r,j}(b),u_{s,j}(b))\right].
\label{eq:profile-residual}
\end{align}
Thus $C^{\mathrm{prof}}$ measures covariance between mean profiles across positions,
while $W$ retains input-dependent covariance after each replica's position-specific means
are removed. We use $W$ to characterize residual organization.

Two properties of $W$ are measured. The first is the self--cross gap that $W$ carries. For any replica matrix $A$, the
self--cross gap is the mean diagonal minus the mean off-diagonal entry,
\begin{equation}
\operatorname{gap}(A)=\E_r[A_{r,r}]-\E_{r\ne s}[A_{r,s}],
\label{eq:matrix-self-cross-gap}
\end{equation}
and the decomposition makes $\operatorname{gap}(Q^{\mathrm{raw}})$ the sum of the gaps of its three
terms. We report $\operatorname{gap}(Q^c)$ and the share $\operatorname{gap}(W)/\operatorname{gap}(Q^c)$
that survives mean-profile subtraction.

The second is the geometry of $W$, which we compare across replicas independently of their
residual magnitudes by normalizing
\begin{equation}
R^W_{r,s}=\frac{W_{r,s}}{\sqrt{W_{r,r}W_{s,s}}},
\label{eq:normalized-profile-residual}
\end{equation}
so that $R^W_{r,r}=1$ and replicas with different residual norms become directly
comparable. We use cosine dissimilarity as the pair distance,
\begin{equation}
d_{r,s}=1-R^W_{r,s}.
\label{eq:residual-pair-distance}
\end{equation}
The distance keeps the sign of $R^W_{r,s}$. \dfnd{Scalar truth alignment has already been removed,
so two replicas with oppositely signed residuals realize different functions rather than one state
and its global flip.} Anticorrelated residuals are far apart. We build a tree by
average-linkage hierarchical clustering, starting from each replica as its own cluster and
repeatedly merging the two clusters whose mean pair distance
$d(C,C')=|C|^{-1}|C'|^{-1}\sum_{r\in C}\sum_{s\in C'}d_{r,s}$ is smallest. The cophenetic distance
$d^{\mathrm{tree}}_{r,s}$ of a pair is the height at which it first belongs to one merged cluster,
and with $\bar d=\E_{r<s}[d_{r,s}]$ the fit is summarized by
\begin{equation}
\begin{aligned}
r_{\mathrm{coph}}&=\operatorname{corr}_{r<s}(d_{r,s},d^{\mathrm{tree}}_{r,s})
{}=\frac{\operatorname{Cov}_{r<s}(d_{r,s},d^{\mathrm{tree}}_{r,s})}
{\sqrt{\operatorname{Cov}_{r<s}(d_{r,s},d_{r,s})\,
\operatorname{Cov}_{r<s}(d^{\mathrm{tree}}_{r,s},d^{\mathrm{tree}}_{r,s})}},\\
R^2_{\mathrm{tree}}&=1-\frac{\sum_{r<s}(d_{r,s}-d^{\mathrm{tree}}_{r,s})^2}
{\sum_{r<s}(d_{r,s}-\bar d)^2}.
\end{aligned}
\label{eq:tree-fit-metrics}
\end{equation}
Sorting each triple's three similarities as $q_1\leq q_2\leq q_3$ and writing
$\delta_{\mathrm{um}}=(q_2-q_1)/(q_3-q_1)$, an exact ultrametric triangle has
$\delta_{\mathrm{um}}=0$, and the fraction of triples with $\delta_{\mathrm{um}}<0.1$ measures how
close the ensemble is to ultrametricity. The cophenetic correlation, the tree fit and the
near-ultrametric fraction $P(\delta_{\mathrm{um}}<0.1)$ all measure how closely the replica
distances can be represented by a tree. \dfnd{These are descriptive tests of finite-ensemble
organization, not by themselves evidence for replica-symmetry breaking, which requires a
thermodynamic hierarchy.}

A further quantity measures how far apart the residuals point rather than how tree-like
their distances are. Since $W_{r,s}$ is an inner product of residual vectors, $R^W_{r,s}$ is the
cosine of the angle between them. We write $P(R^W<0)$ for the obtuse-pair fraction, the share of
pairs more than a right angle apart. 
Every comparison across regimes below uses all output positions.

\subsection{The three learning regimes}
\label{sec:small-data-glass}

We compare memorization endpoints at the fixed update budget with retreat and recovery
endpoints at the fixed epoch of Table~\ref{tab:glass-training-ensembles}. Retreat and recovery are drawn
from the same large-data runs\pcmk{:} each included run reached
$A_{\mathrm{block}}\geq0.8$ at an earlier epoch, and its subsequent trajectory determines the
endpoint class. A retreat endpoint is below $0.5$ at epoch $150$. A recovery endpoint must first
fall below $0.5$ after reaching $0.8$ and then return to $A_{\mathrm{block}}\geq0.8$ by epoch
$150$. Runs that reach high accuracy and never retreat are stable rather than recovered. \dfnd{They and
the runs that end between $0.5$ and $0.8$ enter neither endpoint class, so the two classes do not
exhaust the runs they are drawn from.} Appendix~\ref{app:endpoint-prevalence} reports the prevalence of
each class. We consider the five retreat ensembles and four memorization ensembles
in Table~\ref{tab:glass-training-ensembles}\pcmk{;} pairing each retreat ensemble with its
recovery set gives fourteen sets in total.

\begin{figure}[!htbp]
\centering
\includegraphics[width=\textwidth]{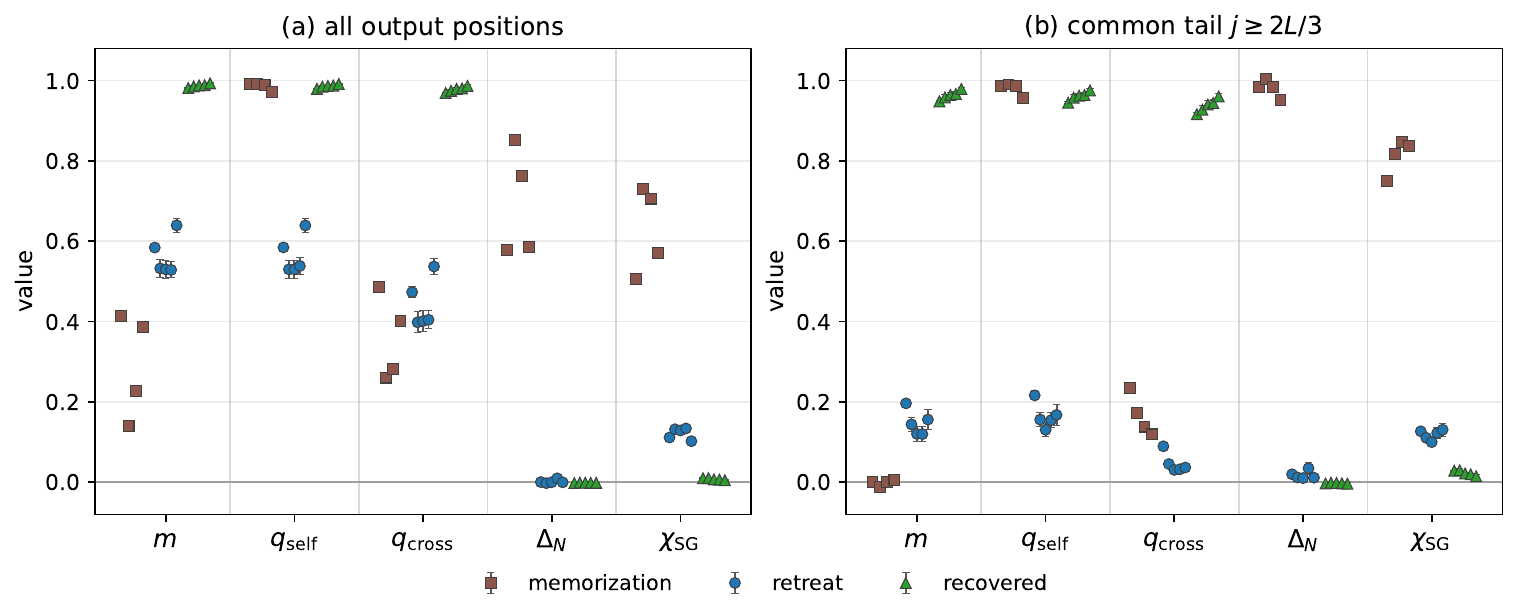}
\caption{Raw overlaps and their gaps of the fourteen replica sets. Panel (a) uses all output positions and
panel (b) uses the common tail $j\geq 2L/3$. Colour and shape give the regime, and within each regime the sets are drawn
left to right in order of $(L,N_{\mathrm{train}})$. The five quantities are defined in
Eqs.~\eqref{eq:binary-truth-overlap}--\eqref{eq:binary-cross-overlap} and
\eqref{eq:Delta-N}--\eqref{eq:chi-SG}. Each error bar shows one standard error. For $m$,
$q_{\mathrm{self}}$ and $\Delta_N$ it is the standard deviation over replicas divided by
$\sqrt{R}$. Pair overlaps that share a replica are correlated. For each replica we first average
its overlap with all other replicas. Writing these averages as $\bar q_r$, the per-replica
self-overlap as $q_{\mathrm{self},r}$, and the sample standard deviation across replicas as $s_R$,
we estimate the standard errors by $2s_R(\bar q_r)/\sqrt{R}$ for $q_{\mathrm{cross}}$ and
$s_R(q_{\mathrm{self},r}-2\bar q_r)/\sqrt{R}$ for $\chi_{\mathrm{SG}}$. The latter retains the
covariance between self- and cross-overlap contributions. Most bars are smaller than the marker.}
\label{fig:order-parameters-sixteen}
\end{figure}

Fig.~\ref{fig:order-parameters-sixteen} first places all the sets on the raw overlaps defined
in Sec.~\ref{sec:binary-functions} together with their two derived gaps. This comparison precedes
the decomposition of Sec.~\ref{subsec:profile-residual-decomposition}. The memorization sets lie
far from $q_{\mathrm{self}}=m$ in both position windows. The retreat and recovery sets both remain
close to it, so $\Delta_N$ does not separate them, while their raw overlaps do. Recovery
has $m$, $q_{\mathrm{self}}$ and $q_{\mathrm{cross}}$ close to one. Retreat has intermediate $m$
and $q_{\mathrm{self}}$, while $q_{\mathrm{cross}}$ is substantially smaller. Consequently
$\chi_{\mathrm{SG}}$ is an order of magnitude larger in retreat.

\paragraph{Retreat.} For the $L=12$, $N_{\mathrm{train}}=1280$ ensemble, the averaged gap of
Eq.~\eqref{eq:matrix-self-cross-gap} decomposes exactly term by term,
\begin{equation}
0.111=\underbrace{0.031}_{\boldsymbol{m}\boldsymbol{m}^\mathsf{T}}+\underbrace{0.080}_{Q^c},
\qquad
0.080=\underbrace{0.018}_{C^{\mathrm{prof}}}+\underbrace{0.062}_{W},
\label{eq:retreat-gap-decomposition}
\end{equation}
so $W$ retains $0.062/0.080=77\%$ of the connected gap. Restricting the evaluation set to the
configurations outside the training set leaves $\operatorname{gap}(Q^c)$ and
$\operatorname{gap}(W)$ unchanged. A percentile bootstrap resamples the 269
replicas with replacement 4000 times and takes the central 95\% of the recomputed gaps. This gives
$[0.102,0.120]$, $[0.074,0.087]$ and $[0.057,0.069]$ for the raw, connected and residual gaps, conditional on the fixed training set,
evaluation set and endpoint selection. The scalar truth alignment removed by
Eq.~\eqref{eq:raw-connected-overlap} is nevertheless substantial. Across the
$\binom{269}{2}=36046$ pairs,
$\operatorname{corr}_{r<s}(Q^{\mathrm{raw}}_{r,s},m_rm_s)=0.966$. What survives is organized by the output
position, concentrating near the learning frontier and beyond, with the last six positions
carrying $71\%$ of the residual gap. Appendix~\ref{app:regime-phenomenology} resolves both
statements by position, and the concentration is what fixes the tail windows of
Table~\ref{tab:glass-training-ensembles}.

\begin{figure}[!htbp]
\centering
\includegraphics[width=\textwidth]{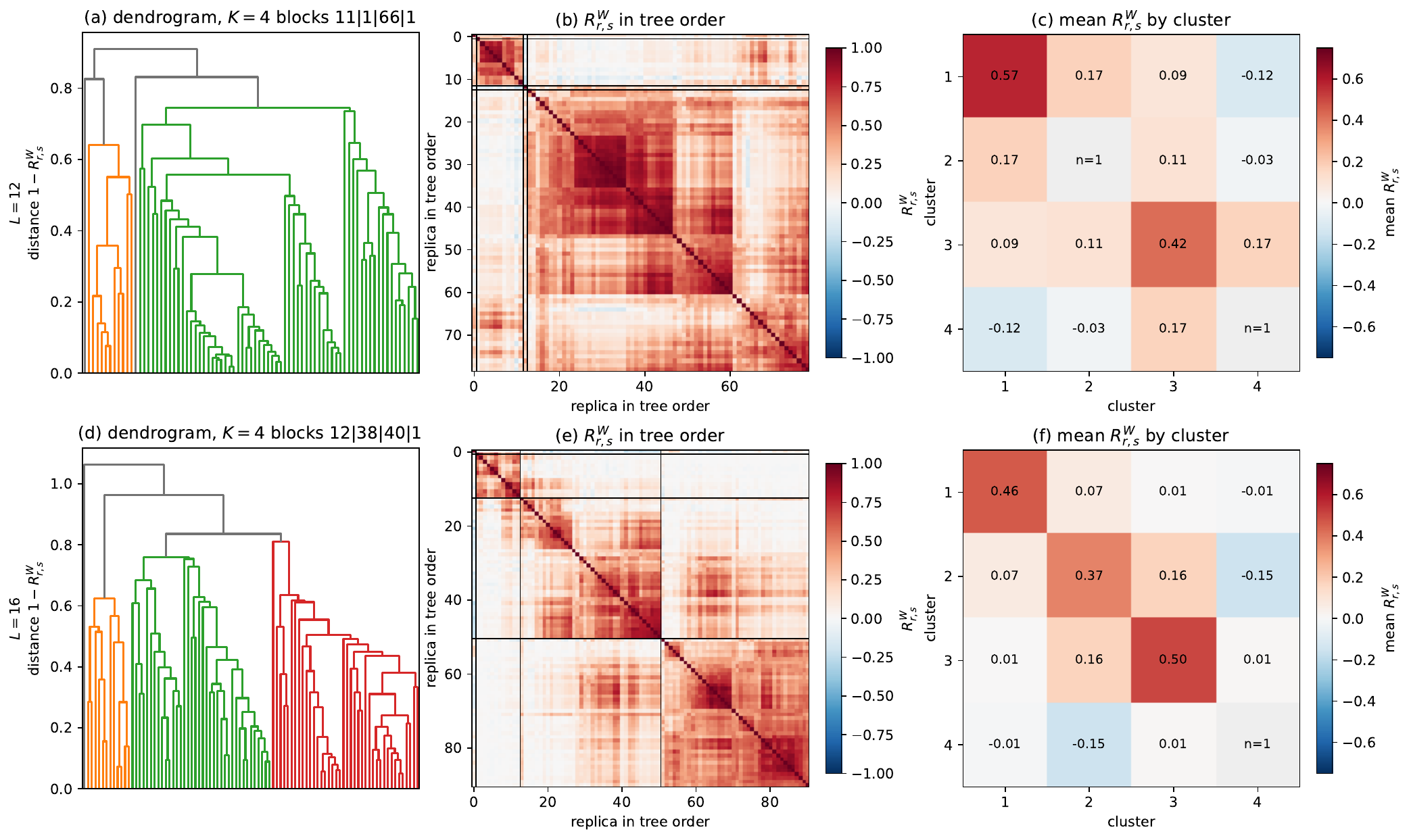}
\caption{Finite-size tree-like organization of the mean-profile-subtracted residual $W$ in the
retreat regime at two string lengths. Panels (a)--(c) use $L=12$, $N_{\mathrm{train}}=1280$ and panels (d)--(f) use $L=16$,
$N_{\mathrm{train}}=2048$. Each displayed set is obtained after applying the bottom-5\% residual-norm filter. At $L=12$ it acts on
the 84 endpoints from the first 400 seeds and leaves 79. At $L=16$ it acts on the full ensemble of
96 and leaves 91. (a), (d) Average-linkage dendrogram for
$d_{r,s}=1-R^W_{r,s}$, with the $K=4$ block sizes given in the panel title. (b), (e)
Normalized residual overlap in tree order, with black lines at the cluster boundaries.
Both panels use the same diverging colour scale centred on zero. (c), (f) Mean overlap for
the $K=4$ cut, where ``n=1'' marks a singleton cluster, which has no within-cluster pair. The
populated clusters show substantially stronger within-cluster than between-cluster similarity.
The two populated clusters of (c) have within-cluster means $0.57$ and $0.42$
against $0.09$ between them.}
\label{fig:retreat-residual-tree}
\end{figure}

The residual geometry is tree-like at finite size
(Fig.~\ref{fig:retreat-residual-tree}). The gap above uses all 269 selected endpoints. The tree
shown for $L=12$ instead uses 84 of these endpoints. The
bottom-$5\%$ residual-norm filter leaves 79. Its illustrative $K=4$ cut has one cluster of $66$ of the $79$ replicas, a second of $11$
and two singletons, and its within-cluster similarity is a few times the between-cluster value.
At $L=16$ the cut is more even, with clusters of $40$, $38$ and $12$ and one singleton. \dfnd{The two
cuts rest on different replica counts and are not comparable across lengths.} Sec.~\ref{subsec:tree-residual-organization}
compares the ensembles at a matched replica count. \dfnd{The value $K=4$
is fixed for display and does not estimate a number of clusters.}
Residual similarity also varies continuously with truth-frontier depth
(Appendix~\ref{app:regime-phenomenology}). \dfnd{The tree alone does not distinguish this continuous
organization from discrete branches.}

Four controls calibrate these statistics on the $(12,1280)$ ensemble, the largest available.
First, the observed cophenetic correlation, tree $R^2$ and near-ultrametric fraction
($0.862$, $0.743$, $0.291$) exceed every one of the $200$ draws of each of two reference
ensembles. The entry-permutation reference symmetrically permutes the off-diagonal entries of $R^W$ and yields
mean values of $(0.20,0.04,0.13)$. The row-centered reference defines each replica's mean similarity as
$\mu_r=\E_{s\neq r}[R^W_{r,s}]$ and independently permutes all entries within each row of
$R^W_{r,s}-(\mu_r+\mu_s)/2$. It averages the permuted matrix with its transpose, adds back
$(\mu_r+\mu_s)/2$, and resets the diagonal to unity. The corresponding mean values are
$(0.50,0.25,0.10)$. \dfnd{These are
matrix-randomization benchmarks\pcmk{:} neither construction guarantees a positive-semidefinite matrix,
and the second does not preserve each row mean exactly, even on average over draws.} Second, tightening the
retreat thresholds from $(0.8,0.5)$ to $(0.9,0.6)$ cuts the retreat count from $269$ to $155$ but
shifts the residual share and $r_{\mathrm{coph}}$ by less than $0.02$. Third, taking epoch 100 rather than
epoch 150 as the observation time shows that only $38\%$ of the runs retreated there are still
retreated at epoch 150. Fourth, the two-time
overlap of a replica's tail output decays within $10$--$25$ epochs to roughly the overlap between
different replicas, $0.27$. Retreat is therefore a state observed during continuing
evolution rather than a frozen endpoint.

The residual share and the tree statistics do not have the same observed dependence on
string length. The residual share falls monotonically\pcmk{:} the three $L=16$ values lie below the
$L=12$ value and the $L=20$ value lies below all three. The tree fit $R^2_{\mathrm{tree}}$ does not follow this ordering.
At the matched replica count of Fig.~\ref{fig:three-regime-dissociation}(d), the $L=20$ set has the
smallest tree fit of the five and falls $0.16$ below the other four, which lie within $0.08$
of each other\pcmk{;} the
near-ultrametric fraction follows neither quantity and peaks at $L=16$. \dfnd{This does not
isolate a size effect\pcmk{:} the experiments fix no scaling path $N_{\mathrm{train}}(L)$,
and changing $N_{\mathrm{train}}$ also changes which endpoints enter the conditioned set.}

\paragraph{Recovery.} These endpoints have completed at least one
retreat--recovery excursion by epoch 150 and come from the same large-data runs as the retreat
sets. Their connected gap is an order of magnitude below the retreat value at every string length,
whereas the residual share remains high and nearly constant across string length. Their
tree fit is larger than the retreat value at every string length
(Fig.~\ref{fig:three-regime-dissociation}(d)). For the illustrative $K=4$ cuts, we use the $L=20$ pair, whose recovery set has $49$
endpoints. Sampling the retreat set down to that count gives mean cluster sizes and standard deviations
$(29.3\pm6.8)/(14.2\pm6.0)/(4.0\pm2.5)/(1.5\pm1.0)$ over $400$ draws, against $44/2/2/1$ for recovery. The recovery cut leaves one
cluster of $44$ and five replicas outside it. The retreat cut has a second cluster of nearly a
third of the replicas. The $L=16$, $N_{\mathrm{train}}=3072$ pair behaves the same way. At its
matched count of $59$ the retreat cut averages $(37.6\pm6.6)/(16.9\pm5.9)/(3.3\pm3.4)/(1.2\pm0.5)$ against $50/5/3/1$ for recovery. The
retreat tree therefore divides its replicas more evenly than the recovery tree.
Sec.~\ref{subsec:tree-residual-organization} extends the comparison to all fourteen sets.

\paragraph{Memorization.} Every replica of the $L=12$, $N_{\mathrm{train}}=64$ ensemble
reaches unit bit accuracy on the shared training set. On the complete held-out set of
$2^{11}-64=1984$ configurations, however, the mean bit accuracy is $0.7070\pm0.0038$ and the block
accuracy is only $0.0624\pm0.0031$. The same $64$ examples determine the rule within
the parity class. Their input matrix has full column rank over the binary field $\mathrm{GF}(2)$, so elimination on it
returns a rule with block accuracy $1$ on all $2048$ configurations. Every fixed training set in
Table~\ref{tab:glass-training-ensembles} has full rank in the same sense. \dfnd{Full rank bounds what the training set determines, not
what gradient descent recovers from it.} Parity is a standard case where algebraic reconstruction
is easy and gradient-based learning is not~\cite{barak2022hidden}. Restricted to the tail, the held-out bit accuracy is
$0.5004\pm0.0016$, consistent with random guessing, while the self-overlap remains
$q_{\mathrm{self}}=0.985$. The replicas are therefore confident and unaligned with the target at large $j$,
giving the $\Delta_N$ signature in Fig.~\ref{fig:order-parameters-sixteen}. The accurate early
positions reflect training coverage rather than generalization, as Appendix~\ref{app:regime-phenomenology}
shows. This regime is temporally persistent over the observation window. Once the replicas interpolate, their within-run tail
overlap remains far above the overlap between different replicas as training continues.

Applying Eq.~\eqref{eq:raw-profile-decomposition} to this ensemble on all output positions
gives a decomposition that contrasts with Eq.~\eqref{eq:retreat-gap-decomposition},
\begin{equation}
0.505=\underbrace{0.006}_{\boldsymbol{m}\boldsymbol{m}^\mathsf{T}}+\underbrace{0.499}_{Q^c},
\qquad
0.499=\underbrace{0.007}_{C^{\mathrm{prof}}}+\underbrace{0.493}_{W}.
\label{eq:memorization-gap-decomposition}
\end{equation}
Scalar truth alignment accounts for $1\%$ of the raw gap and the residual retains $99\%$ of
the connected gap, compared with $77\%$ for retreat. Both conclusions sharpen on the tail, where
these replicas carry no truth alignment,
\begin{equation}
0.751=\underbrace{0.001}_{\boldsymbol{m}\boldsymbol{m}^\mathsf{T}}+\underbrace{0.750}_{Q^c},
\qquad
0.750=\underbrace{0.001}_{C^{\mathrm{prof}}}+\underbrace{0.749}_{W}.
\label{eq:memorization-tail-gap-decomposition}
\end{equation}
There scalar truth alignment accounts for $0.1\%$ and the residual retains $99.9\%$.
Scalar truth alignment also fails to organize the pair pattern here. Across the
$\binom{100}{2}=4950$ pairs, $\operatorname{corr}_{r<s}(Q^{\mathrm{raw}}_{r,s},m_rm_s)=0.608$,
and the four memorization sets span $0.460$ to $0.742$. By contrast, retreat and recovery both stay
above $0.92$. Replica-to-replica variation in overall truth alignment and mean profile shape
therefore explains little of the memorization gap. The remaining input-dependent deviations have
a two-scale geometry, with a near-unit diagonal and a much lower off-diagonal background narrowly
distributed around one value. The off-diagonal entries do not split into groups. Independent random energies produce the same approximately equidistant geometry
among distinct states~\cite{derrida1980random,mezard1987spin}. Appendix~\ref{app:regime-phenomenology}
gives the pair statistics and position profiles.

\subsection{Distinguishing learning regimes by residual geometry}
\label{subsec:tree-residual-organization}

We compare all fourteen sets by their connected gap, residual share and residual
geometry, matching replica counts for the tree statistics. The ensembles are conditioned on
training history and endpoint accuracy.

The residual matrix contains both shared and replica-dependent structure. We write
\begin{equation}
\epsilon_{r,j}(b)=\overline{\epsilon}_j(b)+\xi_{r,j}(b),
\qquad
\overline{\epsilon}_j(b)=\E_r[\epsilon_{r,j}(b)],
\qquad
\E_r[\xi_{r,j}(b)]=0.
\label{eq:shared-replica-residual-decomposition}
\end{equation}
The ensemble-mean residual $\overline{\epsilon}_j(b)$ contributes equally to the mean
diagonal and off-diagonal of $W$, cancelling from $\operatorname{gap}(W)$ but remaining in
$R^W$. Thus the gap measures replica-dependent spread, while residual angles also reflect
common alignment.

\begin{figure}[!htbp]
\centering
\includegraphics[width=\textwidth]{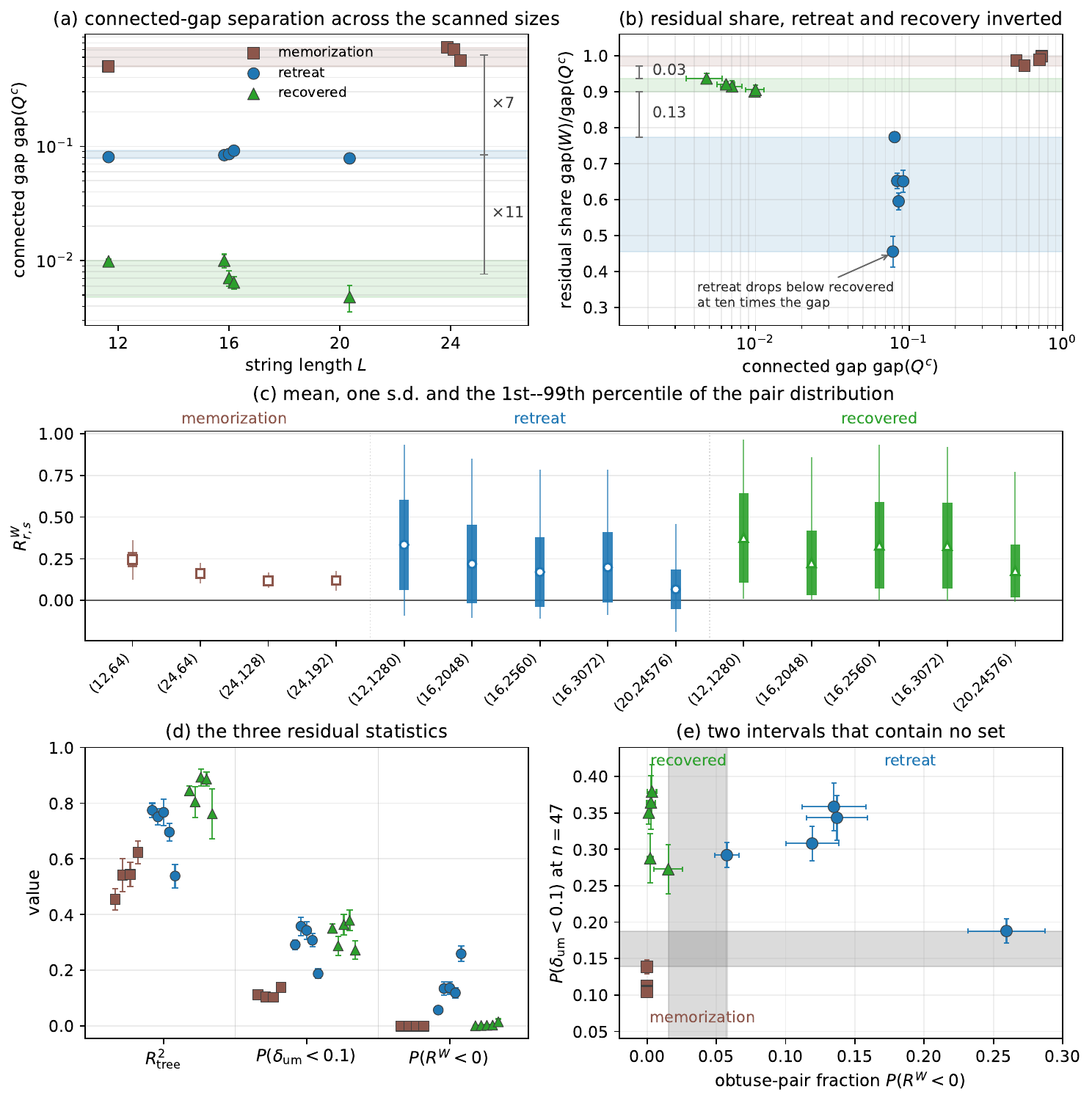}
\caption{Fragmentation amplitude, its decomposition and residual organization for the fourteen
replica sets, on all output positions. Colours and shapes are as in panel (a). The two grey
brackets in (b) span the
interval between adjacent bands and give its width. In (c) the thick bar is the mean
$\pm$ one standard deviation and the thin bar is the 1st to 99th percentile rather than the
extreme, whose length would track the varying pair count. The tree fit
$R^2_{\mathrm{tree}}$ and the near-ultrametric fraction of (d) and (e) are measured on random
subsets of $n=47$ replicas and the obtuse-pair fraction on every replica. Error bars in
(a), (b), (d), and (e) are leave-one-replica jackknife standard errors. Larger ensembles use a
random subset of the deletions.
Grey bands in (e) hold no set.}
\label{fig:three-regime-dissociation}
\end{figure}

Fig.~\ref{fig:three-regime-dissociation}(a) places all fourteen sets on the connected gap. The
three regimes occupy bands separated by approximately an order of magnitude each
throughout the scanned string lengths.

Panel (b) shows that the residual share also separates the three regimes, but orders them
differently\pcmk{:} memorization is highest, followed by recovery and then retreat, with no overlap
between the ranges. Retreat has ten times the connected gap of recovery, yet a smaller residual
share. Fragmentation strength therefore does not determine the relative contribution of
position-profile differences.

Panels (c)--(e) turn to the geometry of the residual. Panel (c) shows the pair
distribution of $R^W$ for each set, and the three regimes are shaped differently. Memorization is
narrow and stays well clear of zero. Retreat is broad and runs through it. Recovery is as broad
as retreat, but the low-overlap end of its distribution remains close to zero. Its first percentile runs from $-0.002$ to $0.015$, whereas the
retreat values run from $-0.185$ to $-0.082$. Each of the three statistics in (d) leaves two of
the three regimes overlapping. The tree fit is built from the whole ensemble and depends
on the replica count, so every set is measured on $400$ independently drawn random subsets of
$n=47$ replicas without replacement and reported as the mean over draws. The near-ultrametric
fraction $P(\delta_{\mathrm{um}}<0.1)$ is measured on the same draws to keep the replica count fixed for this comparison.
It is smaller for every memorization set than for any retreat or
recovery set. Those two overlap in turn. The tree fit $R^2_{\mathrm{tree}}$ is larger for every
recovery set than for any memorization set. The retreat values span both ranges. The
obtuse-pair fraction needs no matched count and uses every replica of each set. It is larger for
every retreat set, at $0.058$ to $0.259$, than for any memorization or recovery set, where it is
$0.016$ or below.

Panel (e) shows that the obtuse-pair and near-ultrametric fractions jointly separate
all three regimes. The five retreat sets have the five largest obtuse-pair fractions among
the fourteen and the four memorization sets the four smallest near-ultrametric fractions, and
neither result uses the other statistic. Three of the four regions so defined hold exactly one
type of set each and the fourth holds none. Of the retreat pairs with $R^W<0$ on the full
evaluation set, $88\%$ to $93\%$ also have $R^W<0$ on each half separately, and the retreat
fractions stay between $0.95$ and $1.05$ times their full-set values. Neither holds for the
recovery sets\pcmk{:} the same share falls to $0.72$ and below, and halving the evaluation set
multiplies their obtuse counts by a factor between zero and eighteen. Their negative $R^W$
entries are less stable under this split-half check. The retreat ensembles therefore show
a larger and more robust anti-correlated component than the memorization and recovery ensembles.

The comparison above uses all output positions. Measured instead on the common window
$j\geq 2L/3$, the three regimes still occupy three non-overlapping ranges of the connected gap,
with memorization highest and recovery lowest. The near-ultrametric fraction now forms three
non-overlapping ranges, with memorization lowest, retreat intermediate and recovery highest, and
the five retreat sets again have the five smallest values of the residual share. On this window the
obtuse-pair fraction separates retreat more sharply still, at $0.18$ to $0.42$ against $0.017$
and below.
Fig.~\ref{fig:small-large-overlap-visibility}(b) places the fourteen sets on this window in the
$P(R^W<0)$--$P(\delta_{\mathrm{um}}<0.1)$ plane. The separation therefore holds for both
choices of averaged positions.

\section{Discussion and outlook}
\label{sec:discussion}

The distinct dynamics and geometry of the failure regimes motivate different
interventions. We summarize the diagnosis and test a learning-rate intervention for retreat.

\subsection{Conclusion}
\label{sec:conclusion}

\begin{figure}[H]
\centering
\includegraphics[width=\textwidth]{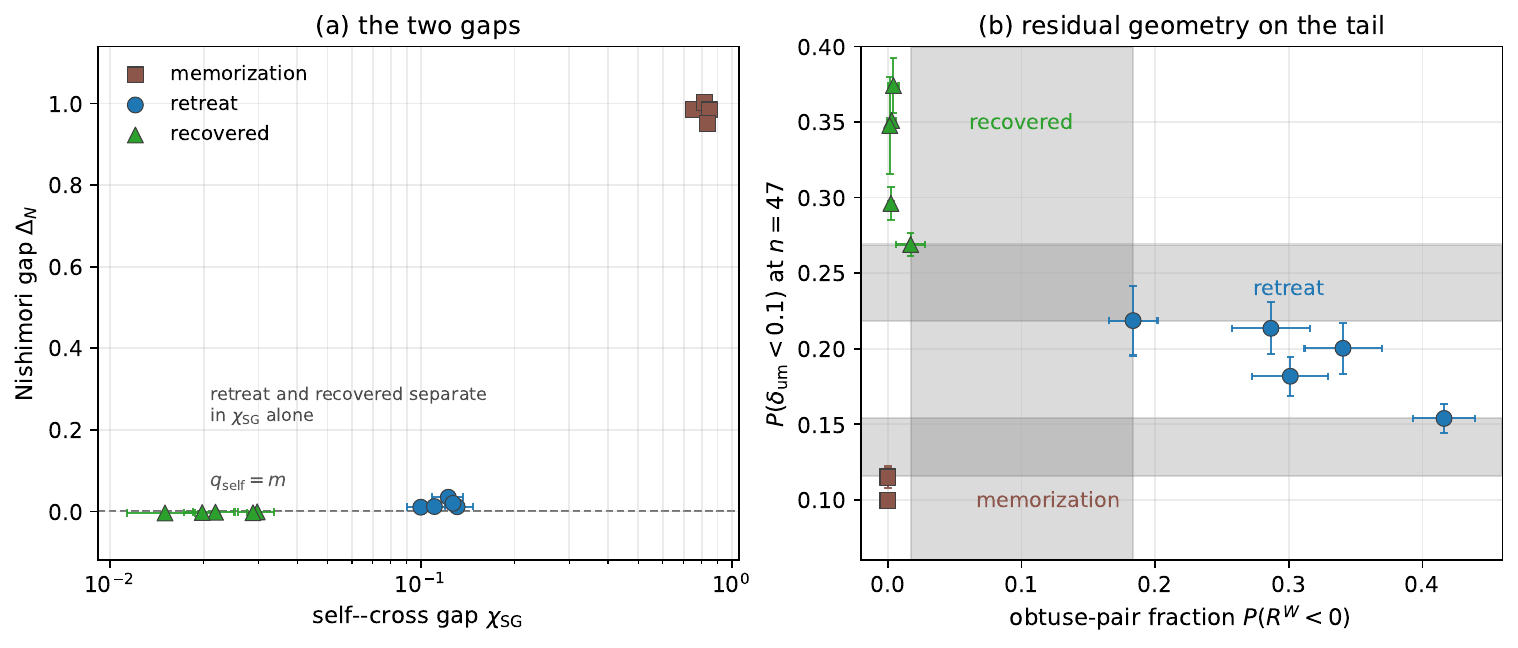}
\caption{The fourteen replica sets in two planes, both on the common tail window
$j\geq 2L/3$. (a) $\chi_{\mathrm{SG}}=q_{\mathrm{self}}-q_{\mathrm{cross}}$ measures how strongly
the ensemble fragments and $\Delta_N=q_{\mathrm{self}}-m$ measures confidence in excess of truth
alignment. (b) The tail-window counterpart of
Fig.~\ref{fig:three-regime-dissociation}(e), with grey bands again marking intervals that hold no
set. On this window the near-ultrametric fraction alone puts the three regimes in three ranges
that do not overlap. Bars in (a) are one standard error, estimated as in
Fig.~\ref{fig:order-parameters-sixteen}, including the joint self- and cross-overlap variation
for $\chi_{\mathrm{SG}}$. They are smaller than the markers vertically
and visible on the logarithmic axis only at small $\chi_{\mathrm{SG}}$. Bars in (b) are the
standard deviation over the $400$ draws vertically and a leave-one-replica jackknife
standard error horizontally. The vertical bars show the spread across random replica
subsets.
The full-output comparison is given in Sec.~\ref{sec:classical-glasses}.}
\label{fig:small-large-overlap-visibility}
\end{figure}

Our function-level analysis separates learning the target from reproducing the same function
across training runs. The fixed training sets in our fragmentation comparison identify the
target within the parity class, yet gradient-based training can fail through persistent
memorization or retreat after generalization. A single trajectory distinguishes these histories.
Replica overlaps additionally reveal which predictions are shared and how the learned functions
differ. The connected gap and residual geometry distinguish memorization, retreat and recovery
on both the full output and the common tail $j\geq 2L/3$
(Figs.~\ref{fig:three-regime-dissociation} and~\ref{fig:small-large-overlap-visibility}).

\subsection{Replica diagnosis and targeted intervention}
\label{subsec:discussion-intervention}

Increasing the training-set size improves learning in the small-data regime.
Retreat instead motivates controlling the optimization dynamics or stopping at generalization onset.

\begin{figure}[!htbp]
\centering
\includegraphics[width=\textwidth]{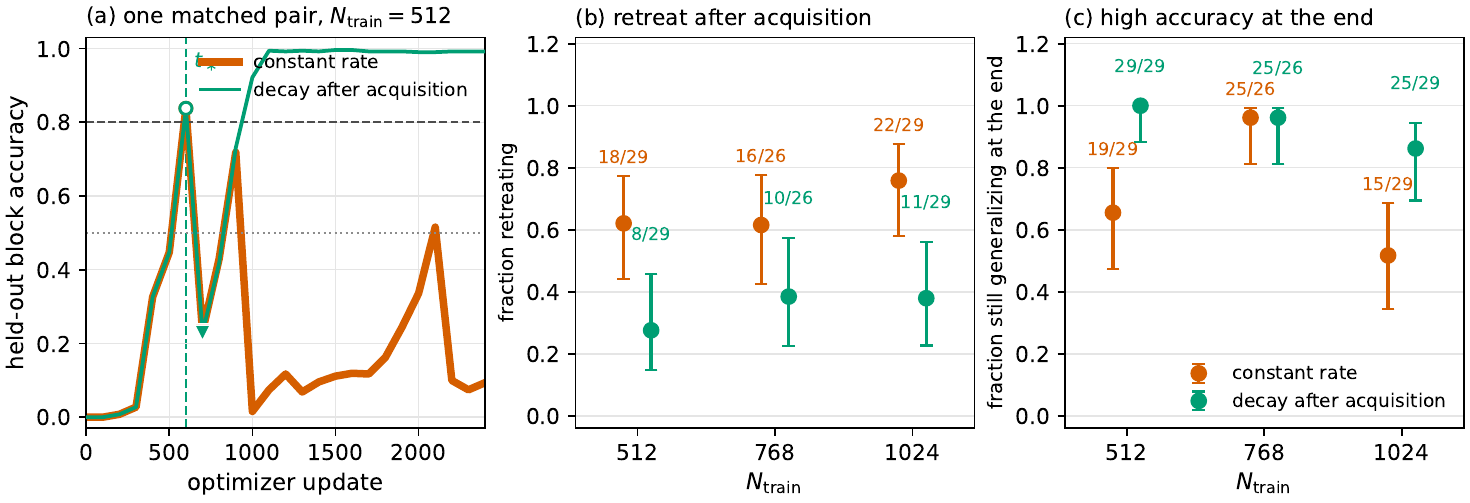}
\caption{Learning-rate intervention for retreat in the $L=12$ Transformer.
(a) Held-out block accuracy for one matched pair at $N_{\mathrm{train}}=512$. The constant-rate
curve is drawn wider beneath the intervention curve to show their overlap through $t_\ast$.
Open circles mark $t_\ast$, and downward triangles mark the first later checkpoint with
$A_{\mathrm{block}}<0.5$. (b) The fraction of runs that retreat. (c) The fraction of runs that
remain in the high-accuracy state at the end of training. Panels (b) and (c) summarize $30$
matched pairs at each size and are conditioned on reaching $A_{\mathrm{block}}\geq0.8$.
Counts are printed above each point. Bars are the Wilson score intervals of
footnote~\ref{fn:wilson} at $95\%$ coverage.
}
\label{fig:cosine-retreat-control}
\end{figure}

We test an optimization-specific intervention for retreat at $L=12$ by reducing the
learning rate after the high-accuracy state is first reached. The constant arm keeps
$\eta(t)=\eta_0$ throughout. The intervention arm holds $\eta_0$ until held-out block accuracy
first reaches $0.8$ at update $t_\ast$, and then follows
$\eta(t)=\eta_{\min}+(\eta_0-\eta_{\min})[1+\cos(\pi(t-t_\ast)/(T-t_\ast))]/2$. Here $\eta(t)$ is
the learning rate at optimizer update $t$, with $\eta_0=2\times10^{-3}$, $\eta_{\min}=0.02\eta_0$,
and $T$ the total update budget. Within each pair, the runs share the data split,
initialization, minibatch stream, optimizer, and update budget. Their trajectories are identical
through $t_\ast$, so subsequent differences arise from the reduced update scale.
Fig.~\ref{fig:cosine-retreat-control}(a) shows one pair at $N=512$. The trajectories coincide
through $t_\ast=600$ and both subsequently fall below $0.5$. The intervention run returns to the
high-accuracy state and ends at $0.99$, whereas the constant-rate run ends at $0.09$.

We compare the two arms in $30$ matched pairs at each of $N=512$, $768$ and $1024$.
Both arms reach the high-accuracy state in the same $84$ of the $90$ pairs because their
trajectories are identical before $t_\ast$. These $84$ matched pairs divide into four outcomes.
Both runs retreat in $24$ pairs. Only the constant-rate run retreats in $32$ pairs, only the
intervention run retreats in $5$ pairs, and neither run retreats in $23$ pairs.

At each training-set size, the retreat fraction is lower under the intervention than
under the constant rate (Fig.~\ref{fig:cosine-retreat-control}(b)). At the end of training,
$79$ of the $84$ intervention runs remain in the high-accuracy state, compared with $59$ under
the constant rate (Fig.~\ref{fig:cosine-retreat-control}(c)). Within the observation window,
reducing the learning rate after generalization therefore lowers the incidence of retreat.

\subsection{Learning string operators beyond one dimension}
\label{subsec:discussion-qec}

The present task can be viewed as a gauge-fixed one-dimensional syndrome-decoding problem. In
two-dimensional surface or toric codes, corrections that differ by contractible stabilizer loops
are physically equivalent, whereas a noncontractible difference changes the logical class. The
replica protocol can therefore be extended to neural decoders that return a correction on the
physical qubits~\cite{torlai2017neural} by defining
overlaps at the level of stabilizer-equivalence classes rather than between raw correction strings.
In this setting, the spatial separation between the two ends of a correction string plays the role
of the output position $j$, which sets the string length in the one-dimensional task. Resolving
$m$, $q_{\mathrm{self}}$, and $q_{\mathrm{cross}}$ as functions of this separation would test
whether fragmentation concentrates on long strings and whether retreat again begins at the largest
length scales.

\subsection{Limitations and future directions}
\label{subsec:discussion-limitations}

The present results describe finite-size behavior rather than asymptotic scaling. They depend on
the system sizes, replica counts, and observation windows studied here. Each fragmented
ensemble is conditioned on one realization of the training data, so it omits the effect of
variation in the dataset. The learning-rate comparison leaves the dependence
on $L$ open.

Future work should extend these ranges, repeat the analysis across training sets,
and compare learning rates, regularization strengths, and optimizers to test whether
retreat is specific to the cyclic late-training dynamics of adaptive optimizers~\cite{thilak2022slingshot}. 
A simpler model whose interpolating functions can be enumerated would separate fragmentation
caused by the admissible function space from that caused by the dynamics. 
Testing whether the regimes also separate in representation space will require
tools that account for the network's permutation and scaling symmetries.

\section*{Code and data availability}

The code that trains the networks, performs the analyses and draws the figures
of this paper is available at \url{https://github.com/HanMaPhy/two_learning_glasses}.
The trained networks and training records it reads are attached to the release
\texttt{data} of that repository.

\section*{Acknowledgments}

We gratefully acknowledge discussions with Zohar Ringel. This work is funded
by the support from the ANR contract associated with the CNRS Chaire
Professeur Junior project ``QOQTIB''. H.M. also thanks École Polytechnique for
support.
I used ChatGPT and Claude Code to assist with coding, generating figures
and checking the grammar of the manuscript, and take full responsibility for the content of this paper.

\appendix

\counterwithin*{figure}{section}
\counterwithin*{table}{section}
\renewcommand{\thefigure}{\thesection\arabic{figure}}
\renewcommand{\theHfigure}{\thesection\arabic{figure}}
\renewcommand{\thetable}{\thesection\arabic{table}}
\renewcommand{\theHtable}{\thesection\arabic{table}}
\setcounter{figure}{0}
\setcounter{table}{0}

\section{Model-level signatures of retreat}
\label{app:retreat-curvature}

To connect retreat in the outputs to changes inside the model, we examine evaluation-loss
gradients, Hessian extrema and the final FFN. These diagnostics complement studies of hidden
progress and run-to-run variation~\cite{barak2022hidden,geiger2020scaling}
by resolving structure lost during retreat.

\subsection{Local evaluation-loss landscape diagnostics}
\label{subsec:hessian-diagnostics}

For a fixed diagnostic set \(D_{\mathrm H}\), we evaluate the mean binary cross-entropy of
Eq.~\eqref{eq:training-bce} and define its parameter Hessian by
\begin{align}
\mathcal L_{D_{\mathrm H}}(\boldsymbol{\vartheta})
&=\frac{1}{|D_{\mathrm H}|L}
\sum_{(b,y)\in D_{\mathrm H}}\sum_{j=0}^{L-1}
\ell_{\mathrm{BCE}}\!\left(z_{\boldsymbol{\vartheta},j}(b),
y_j^{\mathrm{bit}}(b)\right),
\label{eq:hessian-diagnostic-loss}\\
H_{D_{\mathrm H}}(\boldsymbol{\vartheta})
&=\nabla_{\boldsymbol{\vartheta}}^2
\mathcal L_{D_{\mathrm H}}(\boldsymbol{\vartheta}).
\label{eq:hessian-definition}
\end{align}
For a nonzero direction \(v\), its Rayleigh quotient and the two algebraic spectral extrema are
\begin{equation}
\kappa_{D_{\mathrm H}}(v)
=\frac{v^{\mathsf T}H_{D_{\mathrm H}}v}{v^{\mathsf T}v},
\qquad
\lambda_{\min}=\min_{v\ne0}\kappa_{D_{\mathrm H}}(v),
\qquad
\lambda_{\max}=\max_{v\ne0}\kappa_{D_{\mathrm H}}(v).
\label{eq:hessian-rayleigh-extrema}
\end{equation}
We compute Hessian--vector products by second-order automatic differentiation and use
Lanczos/ARPACK iterations to estimate \(\lambda_{\min}\) and \(\lambda_{\max}\) without forming the
full Hessian. The corresponding first-order stationarity diagnostic is
\(\|g\|_2\), where
\(g=\nabla_{\boldsymbol{\vartheta}}\mathcal L_{D_{\mathrm H}}\). Thus
\(\|g\|_2\) is the Euclidean norm over all trainable-parameter components. A negative
\(\lambda_{\min}\) means that the loss curves downward along some direction. A stationary saddle
also requires the gradient norm \(\|g\|_2\) to be small. \dfnd{These raw parameter-space quantities are
parameterization dependent, so we use them only for same-architecture, same-loss comparisons on a
common diagnostic set.}

The fixed-epoch comparison uses 160 \(L=12\),
\(N_{\mathrm{train}}=1280\) replicas at epoch 150, among them the 16 replicas of
Sec.~\ref{subsec:spatial-retreat-recovery}. They share one fixed
training set. The diagnostic set \(D_{\mathrm H}\) is the same for every replica. It consists
of the 1085 domain-wall configurations of the complete space that do not occur in that
training set. Accuracy, loss, gradient norms, and Hessian--vector products are all evaluated
there, so no replica has been trained on any of them.

\subsection{Gradient and curvature at retreat}
\label{subsec:retreat-curvature-results}

We compare the evaluation-loss gradient and Hessian at epoch 150 across 160 replicas run
under the protocol of Sec.~\ref{subsec:spatial-retreat-recovery}. A replica is classified as
``retreat'' if it previously reached $A_{\mathrm{block}}\geq0.8$ but has
$A_{\mathrm{block}}<0.5$ at epoch 150, as ``stable'' if it previously reached the
same threshold and remains above $0.8$, and as ``other'' otherwise. \dfnd{Notice that a stable replica is
not necessarily stationary in parameter space.} This label is set by the endpoint alone.
Table~\ref{tab:endpoint-prevalence} instead reserves ``stable'' for runs that never fall below
$0.5$.

Of the 160 replicas, 131 reach the threshold and 108 of those fall below $0.5$ at some later
epoch. At epoch 150 the three classes hold 50, 66 and 44 replicas. None is stationary with respect to the evaluation loss. The
median $\|g\|_2$ is $0.31$, $0.30$ and $0.47$, and every class has both positive- and
negative-curvature directions. The retreat-minus-stable differences in the medians of
$\|g\|_2$, $\lambda_{\min}$ and $\lambda_{\max}$ are $0.010$, $-1.31$ and $-27.1$, with
$95\%$ bootstrap intervals $[-0.092,0.183]$, $[-5.97,3.04]$ and $[-66.8,14.8]$.\footnote{The
resampling units are drawn with replacement $20{,}000$ times, and the interval spans the central
$95\%$ of the recomputed contrasts. Every interval quoted in this appendix is obtained this way.}
The fixed-epoch comparison therefore resolves no difference between retreat and stable
replicas in these local diagnostics. We next compare changes within individual training runs.
For each retreat we take the first checkpoint with
$A_{\mathrm{block}}<0.5$, hereafter the retreat checkpoint, and the last checkpoint with
$A_{\mathrm{block}}\geq0.8$ before it, which we call the last-high checkpoint. This interval is compared with a nonoverlapping stable interval from
the same seed at the same lag. Checkpoints are saved once per epoch, so the lag is a number of
epochs. Held-out block accuracy remains above $0.8$ throughout the stable
interval. For any diagnostic $X$, we subtract the change from stable start to
stable end from the change between the last-high and retreat checkpoints\pcmk{:}
\begin{equation}
\Delta\!\Delta X
=\left(X_{\mathrm{retreat}}-X_{\mathrm{last\mbox{-}high}}\right)
-\left(X_{\mathrm{stable,end}}-X_{\mathrm{stable,start}}\right).
\label{eq:retreat-difference-in-differences}
\end{equation}
The Hessian analysis uses the first retreat event from each of 12
seeds. They are equally spaced in the sorted seed list and chosen without
reference to gradient or curvature. With the last-high, retreat, stable-start, and stable-end checkpoints,
this gives 48 Hessian evaluations. All checkpoints are scored on one common fixed 1024-configuration
diagnostic set, and the resampling unit for the intervals below is the seed. In this subset,
\(\Delta\!\Delta\lambda_{\min}=-110.3\), with 95\% interval
\([-330.2,90.88]\), and \(\Delta\!\Delta\lambda_{\max}=316.6\), with interval
\([-27.61,761.3]\). The gradient calculation covers all 44 retreat events
found across 24 seeds. Some seeds contribute multiple events, so events are first averaged
within a seed. The
means across seeds are
$\Delta\|g\|_{2,\mathrm{retreat}}=7.110$ and
$\Delta\|g\|_{2,\mathrm{stable}}=0.021$. Their difference is
\(\Delta\!\Delta\|g\|_2=7.089\), with interval \([3.917,11.09]\).
The high-accuracy stable intervals do not sit at exact stationary points, but their
gradient norm shows almost no mean change over the same checkpoint lag.
Retreat is instead marked by a large increase in gradient
norm that is consistent across seeds. Since the intervals for both Hessian contrasts contain zero, the
clearest local parameter-space signature of retreat is this enhanced
nonstationarity, rather than a consistent shift in either Hessian extreme.

\subsection{Target-specific structure in the final FFN}
\label{subsec:ffn-decomposition}

\begin{figure}[H]
\centering
\includegraphics[width=\textwidth]{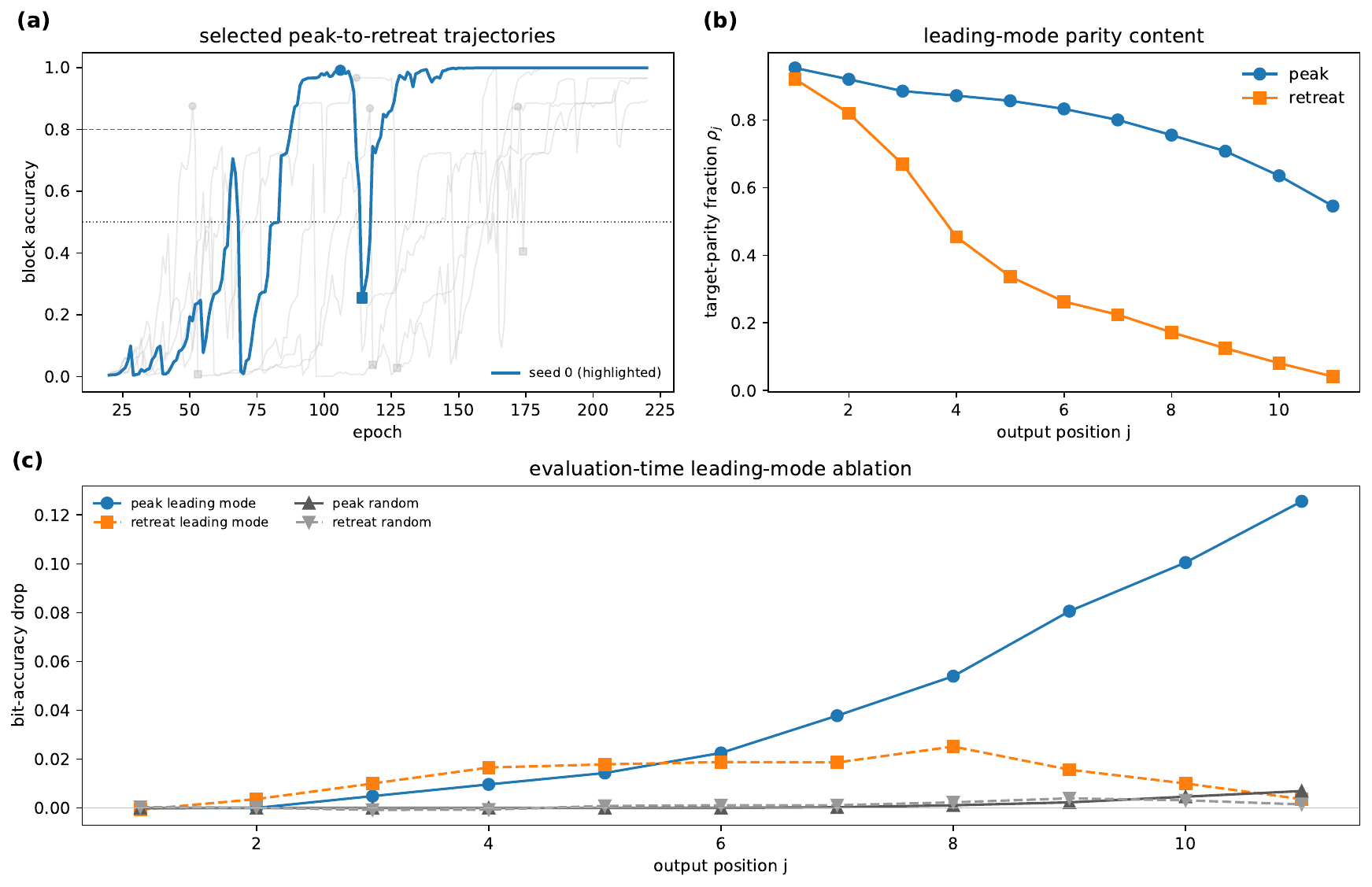}
\caption{Final-layer FFN analysis for 25 peak-to-retreat pairs selected from 30
shared-training-set trajectories. (a) Five representative trajectories, with
seed 0 highlighted in blue and the others shown in gray\pcmk{;} circles mark the
highest-accuracy pre-retreat checkpoint and squares the first subsequent
checkpoint with $A_{\mathrm{block}}<0.5$. (b) The target-parity
fraction $\rho_j$, averaged over the 25 pre-retreat (blue circles) and retreat
(orange squares) checkpoints. (c) Bit-accuracy loss after removing the leading
rank-one component (blue and orange) or a random rank-one component of the
same norm (gray). Panels (b) and (c) show means over the 25 checkpoint pairs. Each
random-control scalar is first averaged over five random realizations at each checkpoint.
All three panels enumerate the complete input space of
\(2^{11}=2048\) distinct domain-wall configurations.}
\label{fig:matched-ffn-transport}
\end{figure}

We next examine the FFN closest to the output head (Fig.~\ref{fig:matched-ffn-transport}). Of
$30$ trajectories, $25$ retreat. For each of them, we compare the checkpoint with the highest block accuracy before
retreat (the peak, which need not coincide with the last-high checkpoint) with the first later checkpoint at which
$A_{\mathrm{block}}<0.5$. Fig.~\ref{fig:matched-ffn-transport}(a) illustrates
this selection for five trajectories, with one highlighted for clarity\pcmk{;}
circles and squares mark the two checkpoints.

We recompute the leading FFN mode at each checkpoint to compare its functional contribution.
Let $h_j(b)\in\mathbb R^{16}$ be the hidden representation at position $j$
entering the FFN of the final Transformer layer, after its attention update.
The FFN computes
\begin{equation}
\phi_j(b)=\operatorname{GELU}\!\left(
W_{\mathrm{up}}\operatorname{LN}(h_j(b))+c_{\mathrm{up}}\right)
\in\mathbb R^{64},
\qquad
h'_j(b)=h_j(b)+W_{\mathrm{down}}\phi_j(b)+c_{\mathrm{down}},
\label{eq:final-ffn-map}
\end{equation}
where $\operatorname{LN}$ denotes layer normalization and
$W_{\mathrm{down}}\in\mathbb R^{16\times64}$ maps the expanded FFN
activation back to the 16-dimensional hidden representation passed to the
final normalization and output layer. We analyze this matrix through its
singular-value decomposition,
\begin{equation}
W_{\mathrm{down}}=\sum_a \sigma_a u_a v_a^{\mathsf T},
\qquad
W_{\mathrm{down}}^{(-1)}
=W_{\mathrm{down}}-\sigma_1u_1v_1^{\mathsf T}.
\label{eq:ffn-leading-mode}
\end{equation}
With $\sigma_1$ the largest singular value, the leading component of the FFN update is
$\sigma_1u_1a_j(b)$, where $a_j(b)=v_1^{\mathsf T}\phi_j(b)$. Replacing
$W_{\mathrm{down}}$ by $W_{\mathrm{down}}^{(-1)}$ is the leading-mode
ablation used below.

We evaluate $a_j(b)$ on all $2^D=2^{11}=2048$ domain-wall configurations and
expand its centered value in the Walsh basis,
\begin{equation}
a_j(b)-\langle a_j\rangle_b
=\sum_{S\ne\varnothing}\widehat a_{j,S}\,\chi_S(b),
\qquad
\chi_S(b)=(-1)^{\sum_{k\in S}b_k},
\label{eq:ffn-walsh-decomposition}
\end{equation}
with
\begin{equation}
\langle a_j\rangle_b=2^{-D}\sum_b a_j(b),
\qquad
\widehat a_{j,S}=2^{-D}\sum_b
\bigl(a_j(b)-\langle a_j\rangle_b\bigr)\chi_S(b).
\label{eq:ffn-walsh-coefficients}
\end{equation}
The target at position $j$ is the Walsh function
$\chi_{[1,j]}$, as defined in Eq.~\eqref{eq:signed-target-character}. We
therefore measure
\begin{equation}
\rho_j=\frac{|\widehat a_{j,[1,j]}|^2}
{\sum_{S\ne\varnothing}|\widehat a_{j,S}|^2}.
\label{eq:ffn-target-parity-fraction}
\end{equation}
Thus $\rho_j$ is the fraction of the nonconstant variation in the leading FFN
activation with exactly the parity dependence required at position $j$.
A related construction tracks grokking~\cite{gromov2023grokking}, with the
Fourier basis of modular arithmetic replaced here by the Walsh basis of parity. \dfnd{What differs is the
use\pcmk{:} $\rho_j$ is resolved by output position and is read across a retreat rather than across
acquisition.}

In Fig.~\ref{fig:matched-ffn-transport}(b), the blue circles show $\rho_j$ at
the pre-retreat checkpoints and the orange squares show it after retreat. The
two curves separate increasingly at late output positions. Across the 25
pairs, the mean falls from 0.797 to 0.374 over all nonconstant positions and from
0.713 to 0.151 for $j\geq6$. In Fig.~\ref{fig:matched-ffn-transport}(c), we
replace $W_{\mathrm{down}}$ by
$W_{\mathrm{down}}-\sigma_1u_1v_1^{\mathsf T}$ at evaluation time, keep all
other parameters fixed, and plot the resulting decrease in bit accuracy. The
blue circles show a growing accuracy loss toward late positions
at the pre-retreat checkpoint, whereas the orange squares show a much smaller
loss after retreat. The two gray triangle curves average the ablation effect of five random
rank-one matrices of the same norm $\sigma_1$ at each checkpoint and remain near zero.
The leading mode thus loses target-parity content and functional influence most strongly
at the late positions affected by retreat.

\section{Relation between bit and block accuracy}
\label{app:accuracy-correlation-statistics}

To quantify how block accuracy amplifies changes in bit accuracy, we compare their
trajectories and test an independent-position estimate.

\begin{figure}[!htbp]
\centering
\includegraphics[width=\textwidth]{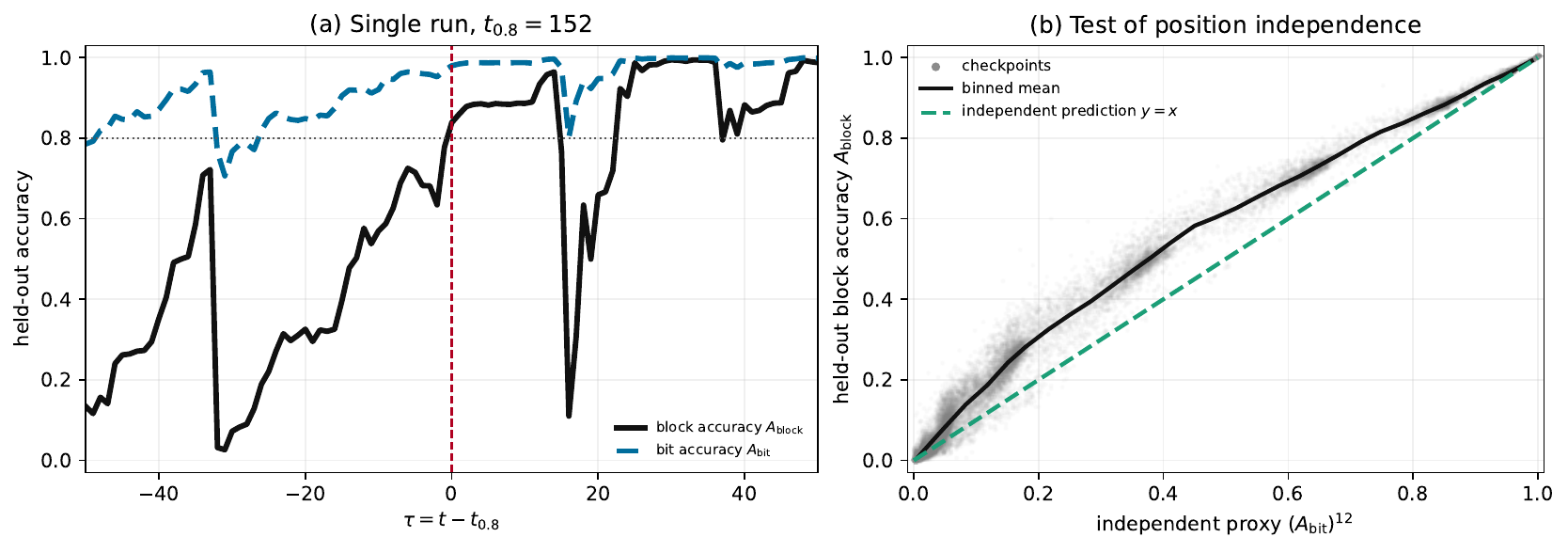}
\caption{Bit and block accuracy for Transformers trained with independently sampled
fixed training sets. (a) Held-out bit and block accuracy for one $L=12$, $N=1280$ run,
shown around its first $A_{\mathrm{block}}=0.8$ crossing at epoch 152. Bit accuracy also changes in
jumps, but less strongly than block accuracy. (b) Measured block accuracy versus the equal-position
independent estimate $(A_{\mathrm{bit}})^{12}$ for the $L=12$
training-set-size scan. The scan comprises 310 independently trained Transformer runs in total,
spread over 21 values of $N$. One checkpoint per epoch for epochs 1--120 gives
$310\times120=37{,}200$ points. Gray points are individual checkpoints, the black curve is
their binned mean, and the green dashed line is the independent prediction.}
\label{fig:transition-observable-overview}
\end{figure}

The two panels of Fig.~\ref{fig:transition-observable-overview} together show that requiring
every position to be correct sharpens a change already visible in bit accuracy. Panel (b) pairs
the measured block accuracy $Y$ of every checkpoint with its equal-position independent estimate
$X=(A_{\mathrm{bit}})^{12}$ at the same epoch. For $M$ such pairs the Pearson correlation is
\begin{equation}
r=\frac{\sum_{i=1}^{M}\left(X_i-\bar X\right)\left(Y_i-\bar Y\right)}
{\sqrt{\sum_{i=1}^{M}\left(X_i-\bar X\right)^2\sum_{i=1}^{M}\left(Y_i-\bar Y\right)^2}},
\label{eq:pearson-correlation}
\end{equation}
where a bar denotes the mean over the pairs. Pooled over all $M=37{,}200$ checkpoints it is
$r=0.989$. The deviation of a single checkpoint is $\Delta=Y-X$, whose mean signed and absolute
values are $\overline{\Delta}=0.033$ and $\overline{|\Delta|}=0.035$. Centering $X$ and $Y$ across
replicas at every fixed training-set size and epoch removes the systematic variation they share.
The concatenated residuals still give $r=0.983$.

\section{Regime phenomenology}
\label{app:regime-phenomenology}

The geometric comparisons in Sec.~\ref{sec:classical-glasses} depend on endpoint selection
and position averaging. Here we report class prevalence and resolve the fragmentation and
generalization failure by position.

\subsection{Prevalence of the endpoint classes}
\label{app:endpoint-prevalence}

The endpoint classes of Sec.~\ref{sec:small-data-glass} divide an ensemble as shown below for
the $(12,1280)$ runs.

\begin{center}
\begin{forest}
  for tree={
    grow'=0, folder, anchor=west, child anchor=west, align=left,
    font=\small, edge={gray}, l sep=5mm, s sep=1.2mm, inner ysep=1.6pt,
  }
  [{$1200$ trained runs}
    [{$255$ never reach $A_{\mathrm{block}}=0.8$ within 150 epochs \textit{(never)}}]
    [{$945$ reach $0.8$}
      [{$145$ never fall below $0.5$ afterwards \textit{(stable)}}]
      [{$800$ fall below $0.5$ after first reaching $0.8$}
        [{$269$ below $0.5$ at epoch 150 \textit{(retreat)}}]
        [{$416$ at or above $0.8$ at epoch 150 \textit{(recovery)}}]
        [{$115$ between $0.5$ and $0.8$ at epoch 150 \textit{(neither)}}]
      ]
    ]
  ]
\end{forest}
\end{center}

Table~\ref{tab:endpoint-prevalence} gives these five shares for each of the five large-data
ensembles. They are taken over all trained runs and add to one, so the $(12,1280)$ column entries
are the five leaves above divided by $1200$. ``Returning'' has a different denominator. It takes
the $800$ runs that fall below $0.5$ and counts those that regain $0.8$ at any point before epoch
150, whether or not they hold it, which is $531$ here.

The never-reaching share follows $N_{\mathrm{train}}$ relative to the crossover scale $N_c$
rather than $L$. The two ensembles at the same ratio $1.2$ but different string length sit at
$0.21$ and $0.25$. The stable share instead falls with string length and reaches zero at
$(20,24576)$, where every run that reaches $0.8$ later falls below $0.5$. The
retreat share grows with string length from $0.22$ to $0.50$, while the returning fraction does
not fall. More replicas are therefore caught in retreat at epoch 150 at the
longer strings, and not because each retreat has become harder to leave.

\begin{table}[!htbp]
\centering
\small
\begin{tabular}{@{}cccccccc@{}}
\toprule
& & \multicolumn{5}{c}{share of trained runs} & \\
\cmidrule(lr){3-7}
$(L,N_{\mathrm{train}})$ & $N_{\mathrm{train}}/N_c$ & never & stable & retreat & recovery
& neither & returning \\ \midrule
$(12,1280)$  & $1.2$ & $0.21$ & $0.12$ & $0.22$ & $0.35$ & $0.10$ & $0.66$ \\
$(16,2048)$  & $0.8$ & $0.46$ & $0.05$ & $0.29$ & $0.16$ & $0.04$ & $0.61$ \\
$(16,2560)$  & $1.0$ & $0.42$ & $0.06$ & $0.30$ & $0.16$ & $0.06$ & $0.65$ \\
$(16,3072)$  & $1.2$ & $0.25$ & $0.06$ & $0.41$ & $0.23$ & $0.05$ & $0.67$ \\
$(20,24576)$ & $2.6$ & $0.16$ & $0.00$ & $0.50$ & $0.31$ & $0.03$ & $0.83$ \\ \bottomrule
\end{tabular}
\caption{Endpoint classes in the five large-data ensembles. $N_c$ is the fitted crossover
scale of Eq.~\eqref{eq:learning-probability-logistic}. The five middle columns are fractions of
all trained runs and add to one. ``Never'' is the share that never reaches
$A_{\mathrm{block}}=0.8$ within 150 epochs, and ``stable'' the share that reaches it and never
falls below $0.5$ afterwards. ``Retreat'', ``recovery'' and ``neither'' split the remaining runs
by their block accuracy at epoch 150, following Sec.~\ref{sec:small-data-glass}. ``Returning''
has a different denominator. It is the fraction of the runs that fall below $0.5$ which regain
$0.8$ at any point before epoch 150.}
\label{tab:endpoint-prevalence}
\end{table}

\subsection{The moving truth frontier of the retreat ensembles}

\begin{figure}[!htbp]
\centering
\includegraphics[width=\textwidth]{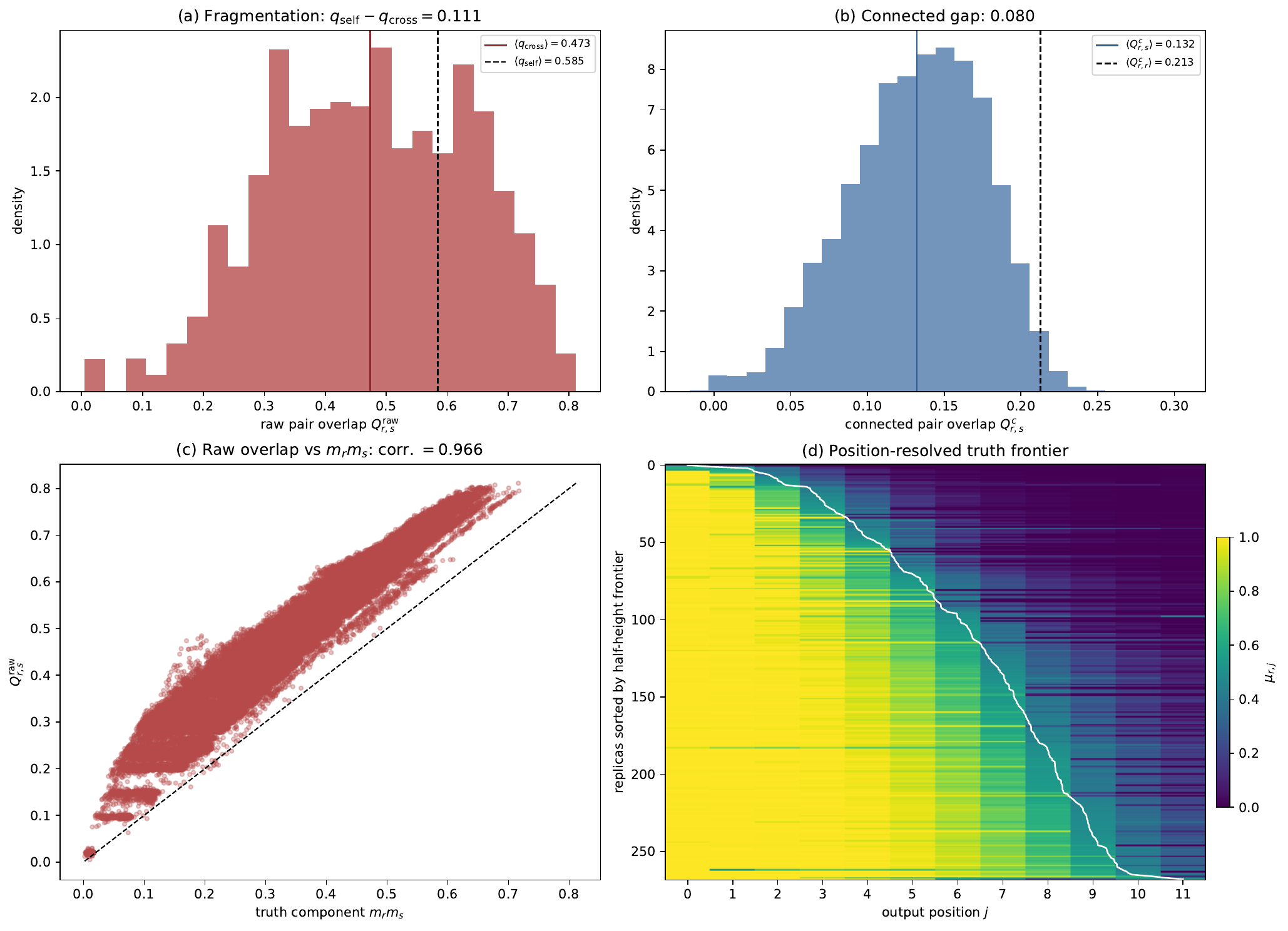}
\caption{Raw fragmentation and truth-frontier order for the 269 selected retreat endpoints on one
shared fixed training set. (a) Raw pair overlaps and the self--cross gap. (b) The gap that remains
after subtracting $m_rm_s$. (c) The raw overlap is strongly organized by the scalar truth
component $m_rm_s$. (d) Position-resolved profiles sorted by the position \(j\) at which each profile crosses one half (white curve) show
an ordered small-$j$ region, a broad moving frontier, and weak truth alignment at large $j$.}
\label{fig:retreat-overlap-structure}
\end{figure}

\begin{figure}[!htbp]
\centering
\includegraphics[width=0.88\textwidth]{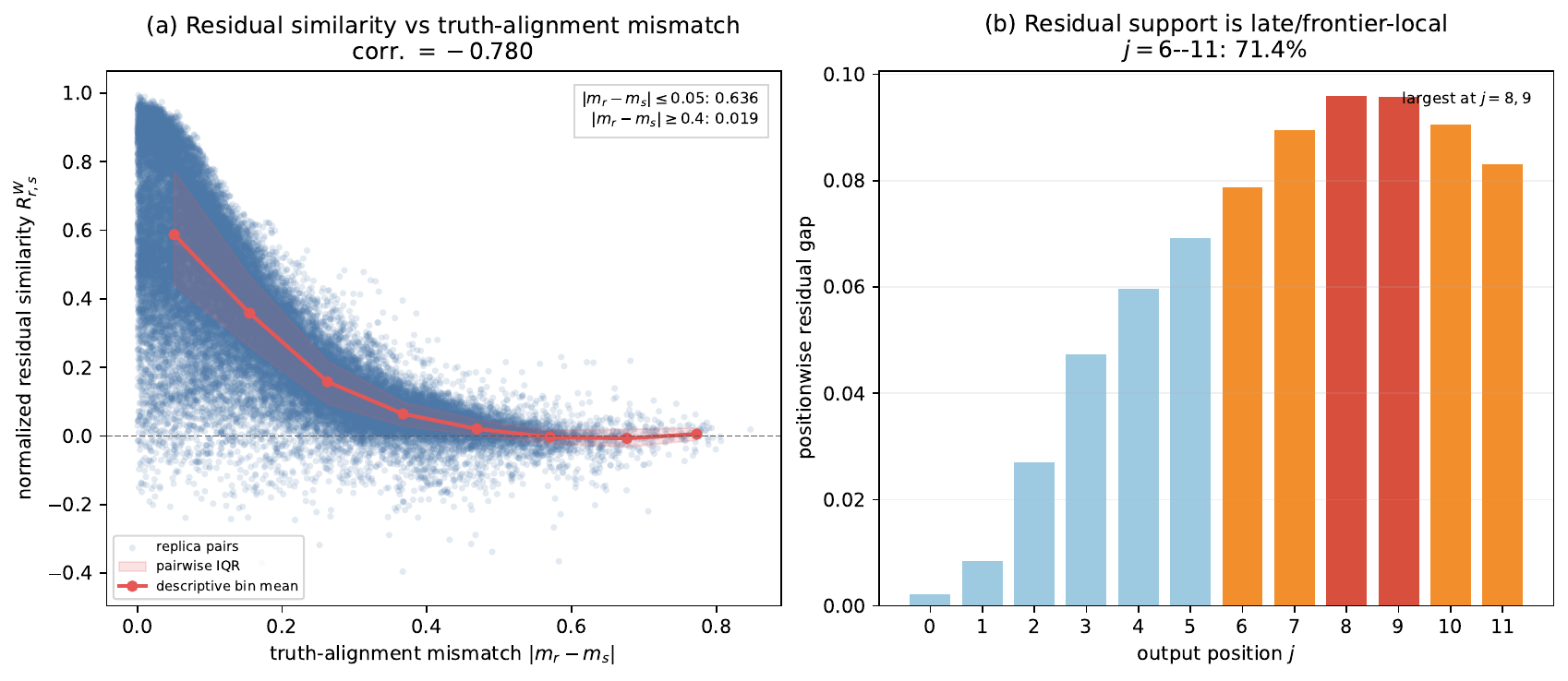}
\caption{(a) Normalized residual similarity
decreases continuously as the difference in overall truth alignment increases\pcmk{;} the heavy curve shows the mean within each interval of $|m_r-m_s|$. 
The band spans the 25th to 75th percentiles of the pair values. (b) The positionwise residual gaps $g_j$, whose average is
$\operatorname{gap}(W)$, are concentrated at late and frontier positions. The residual is therefore structured by the output position even after
each replica's mean position profile has been removed.}
\label{fig:retreat-frontier-residual}
\end{figure}

The truth-alignment profiles $\mu_{r,j}$ carry reproducible truth alignment at small $j$, a
finite-width transition, and weak truth alignment at large $j$. For these frontier-shaped profiles,
a larger $m_r$ corresponds to a frontier extending to larger $j$.
Fig.~\ref{fig:retreat-overlap-structure}(d) orders the replicas by the position at which their
profile crosses one half. The white curve sweeps from the first output positions to the last, so
frontier depth forms a continuous spread rather than a few discrete levels. This spread appears in
the pair overlaps through the scalar component $m_rm_s$.
Fig.~\ref{fig:retreat-overlap-structure}(c) plots $Q^{\mathrm{raw}}_{r,s}$ against $m_rm_s$ for
all $36{,}046$ distinct replica pairs and gives the correlation $0.966$ reported in
Sec.~\ref{sec:small-data-glass}. Panels (a) and (b) show the distributions before and after
subtracting $m_rm_s$. The raw overlaps span values from near zero to $0.8$, while the connected
overlaps form a single narrower mode. The self--cross gap remains $0.080$ after the subtraction,
compared with the raw gap $0.111$. Scalar truth alignment therefore organizes much of the
pair-to-pair variation, while most of the gap remains in the connected overlap.
Residual similarity falls continuously with mismatch in overall truth alignment,
$\operatorname{corr}_{r<s}(R^W_{r,s},|m_r-m_s|)=-0.780$, so pairs at similar frontier depth remain
the most alike even after the scalar has been subtracted
(Fig.~\ref{fig:retreat-frontier-residual}(a)).

To resolve the residual gap by position, let
$K^{(j)}_{r,s}=\E_b[\epsilon_{r,j}(b)\epsilon_{s,j}(b)]$ and $g_j=\operatorname{gap}(K^{(j)})$.
Note that $\operatorname{gap}(W)=\E_j[g_j]$. We obtain positionwise gaps that are near zero at
the early positions where the replicas agree, rise through the frontier and peak at $j=8$ and
$9$, with $\sum_{j=6}^{11}g_j/\sum_{j=0}^{11}g_j=0.714$
(Fig.~\ref{fig:retreat-frontier-residual}(b)). This peak sits where the frontier distribution of
Fig.~\ref{fig:retreat-overlap-structure}(d) is densest, with more than a third of the replicas
crossing one half at $j=8$ or $9$.

\subsection{Position-resolved generalization failure of the memorization ensembles}
\label{subsec:small-data-generalization}

To locate the failure by output position, let $c_j(b)=(b_1,\ldots,b_j)$ be the domain-wall
configuration that determines the output $y_j$. A value of $c_j$ is covered if it occurs in at
least one training configuration, and $C_j$ counts the covered values among the $2^j$
possibilities. Each value of $c_j$ extends to $2^{11-j}$ complete configurations, so the covered
and uncovered groups contain $C_j2^{11-j}-64$ and $(2^j-C_j)2^{11-j}$ held-out configurations. In
this training-set realization all values of $c_j$ are covered through $j=4$ and incomplete coverage
begins at $j=5$. \dfnd{For values absent from training, $A_j^{\mathrm{uncovered}}$ remains near $0.5$ at
every $j\geq5$, so the accuracy at $j\leq4$ does not establish generalization to unseen
first-$j$ input configurations\pcmk{:} every such configuration is already represented in the training set
(Table~\ref{tab:small-data-domain-wall-coverage}).}

\begin{table}[!htbp]
\centering
\small
\begin{tabular}{cccc}
\toprule
$j$ & Covered first-$j$ configurations $C_j/2^j$ & $A_j^{\mathrm{covered}}$ & $A_j^{\mathrm{uncovered}}$ \\ \midrule
1  & $2/2$     & $0.996\pm0.001$ & --- \\
2  & $4/4$     & $0.945\pm0.004$ & --- \\
3  & $8/8$     & $0.884\pm0.006$ & --- \\
4  & $16/16$   & $0.778\pm0.010$ & --- \\
5  & $27/32$   & $0.763\pm0.009$ & $0.487\pm0.019$ \\
6  & $42/64$   & $0.659\pm0.007$ & $0.509\pm0.013$ \\
7  & $49/128$  & $0.696\pm0.006$ & $0.476\pm0.007$ \\
8  & $59/256$  & $0.660\pm0.005$ & $0.481\pm0.004$ \\
9  & $62/512$  & $0.664\pm0.006$ & $0.485\pm0.002$ \\
10 & $63/1024$ & $0.450\pm0.009$ & $0.492\pm0.001$ \\
11 & $64/2048$ & ---             & $0.492\pm0.001$ \\ \bottomrule
\end{tabular}
\caption{Accuracy grouped by whether $c_j$ occurs in training, evaluated on held-out configurations in the
$L=12$, $N_{\mathrm{train}}=64$ ensemble. The trivial constant output at $j=0$ is omitted. A dash
indicates that the corresponding subset is empty\pcmk{:} all possible values of $c_j$ are covered for
$j\leq4$, while no held-out configuration contains a training-covered value of $c_{11}$.}
\label{tab:small-data-domain-wall-coverage}
\end{table}

\subsection{Approximately equidistant residual geometry of the memorization ensembles}
\label{subsec:small-data-rem-geometry}

\begin{figure}[!htbp]
\centering
\includegraphics[width=\textwidth]{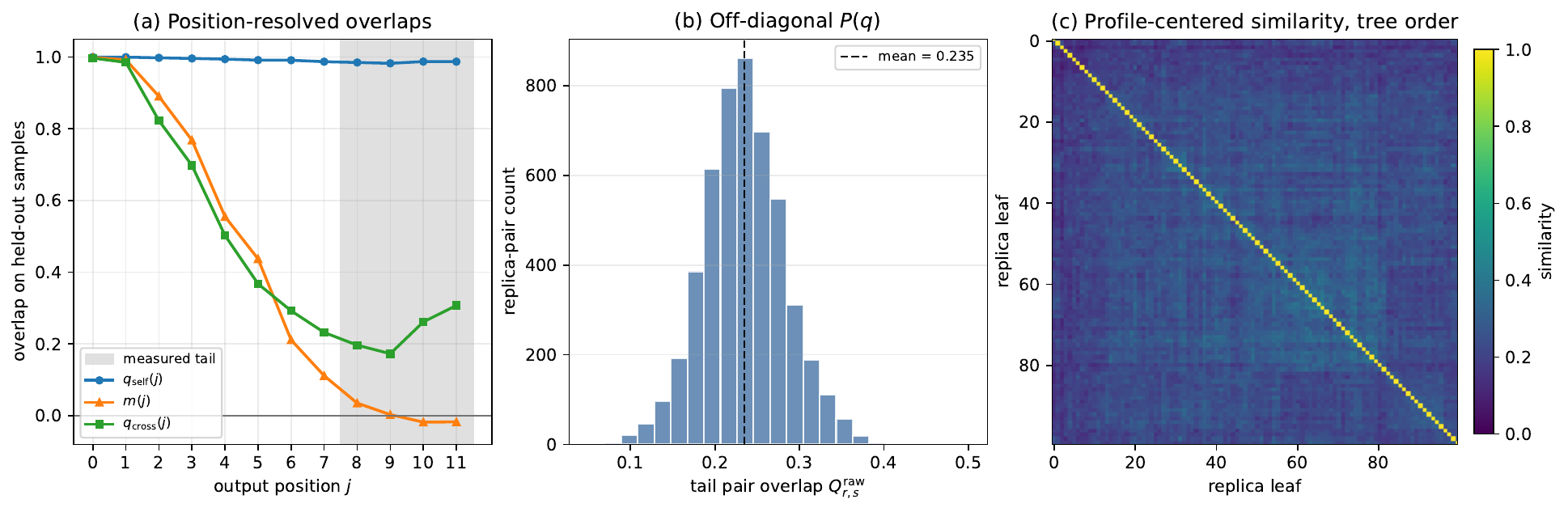}
\caption{Three-stage small-data diagnosis for $L=12$, $N_{\mathrm{train}}=64$, 100 replicas, and
1984 held-out configurations. (a) Raw position-resolved truth, self-, and cross-replica overlaps\pcmk{;} shading
marks the $j=8,\ldots,11$ tail. (b) The 4950 off-diagonal raw tail overlaps. (c) Tail-restricted $R^W$ in tree order after subtracting scalar truth alignment and the
mean position-profile shape. The common $[0,1]$ scale shows weak residual contrast off the
diagonal.}
\label{fig:small-data-overlap}
\end{figure}

In Fig.~\ref{fig:small-data-overlap}(a), memorizing replicas remain confident at every
position while truth alignment and cross-replica agreement fall with algebraic degree.
Restricting all averages to the complete held-out set and
$\mathcal J_{\mathrm{tail}}$, the $\binom{100}{2}=4950$ off-diagonal entries of
$Q^{\mathrm{raw}}$ for the $L=12$, $N_{\mathrm{train}}=64$ ensemble range from $0.066$ to $0.442$
with none above $0.7$, against $q_{\mathrm{self}}=0.985$
(Fig.~\ref{fig:small-data-overlap}(b)). No two replicas in this ensemble realize the
same tail function. The remaining weak off-diagonal modulation prevents strict exchangeability.
It nevertheless does not group the replicas (Fig.~\ref{fig:small-data-overlap}(c)). The
near-ultrametric fractions are reported in Table~\ref{tab:small-data-geometry}. \dfnd{We use the
approximately equidistant geometry as a descriptive reference for the retreat ensembles.}

\begin{table}[!htbp]
\centering
\small
\begin{tabular}{@{}lccccc@{}}
\toprule
Ensemble & $q_{\mathrm{self}}$ & $q_{\mathrm{cross}}$
& Pair s.d. & $R^2_{\mathrm{tree}}$ & $P(\delta_{\mathrm{um}}<0.1)$ \\ \midrule
$L=12,\ N=64$  & 0.985 & 0.235 & 0.048 & 0.512 & 0.116 \\
$L=24,\ N=64$  & 0.990 & 0.178 & 0.033 & 0.480 & 0.101 \\
$L=24,\ N=128$ & 0.986 & 0.150 & 0.025 & 0.525 & 0.098 \\
$L=24,\ N=192$ & 0.957 & 0.126 & 0.024 & 0.544 & 0.111 \\ \bottomrule
\end{tabular}
\caption{Tail-window diagnostics for the four memorization ensembles of
Table~\ref{tab:glass-training-ensembles}. Every entry uses the tabulated tail window and the
evaluation configurations listed there. For $(12,64)$ these are the complete held-out set and
for the $L=24$ rows they are $4000$ fixed draws. Pair s.d. is the standard deviation of the off-diagonal entries of
$Q^{\mathrm{raw}}$. The tree fit uses the full replica ensemble, and the near-ultrametric
fraction uses all its triples, following Sec.~\ref{subsec:profile-residual-decomposition}.
Three uncorrelated distances give $P(\delta_{\mathrm{um}}<0.1)=0.1$, and these ensembles sit at
that value on their tail.
}
\label{tab:small-data-geometry}
\end{table}
\bibliographystyle{unsrt}
\bibliography{references}

\end{document}